\documentclass[acmsmall]{acmart}

\usepackage{amssymb,amsmath,amsthm}
\usepackage{latexsym}
\usepackage{graphicx}
\usepackage{textcomp}
\usepackage{verbatim} 
\usepackage{xspace} 
\usepackage{mathpartir}
\usepackage{cancel}
\usepackage{tikz}
\usetikzlibrary{arrows,automata,calc,decorations.text,shapes}
\usepackage{xxcolor}
\usepackage{multirow}
\usepackage{stmaryrd}
\usepackage{bbm}

\usepackage{booktabs}   
\usepackage{subcaption} 

\usepackage{skull}

\newcounter{ToDos}
\newcounter{WarnCounts}
\newcommand{\decorateWC}{
  \stepcounter{WarnCounts}
  \marginpar{\textcolor{red}{$\skull\ \theWarnCounts$}}}

\newcommand{\act}[1]{\mathsf{#1}}
\newcommand{\hmapsto}{\Mapsto}
\newcommand{\theheap}{\sigma}

\newcommand{\thelock}{\mu}
\newcommand{\thehist}{\tau}
\newcommand{\ldot}{\mathord{.}\,}
\newcommand{\eqdef}{\mathrel{\:\widehat{=}\:}}
\newcommand{\eqbool}{\mathbin{=}}

\newcommand{\vrf}{\mathsf{vrf}}
\newcommand{\wpeq}[2]{\vrf\ #1\ #2}
\newcommand{\wpeqs}[3]{\vrf\ #1\ #2\ #3}

\newcommand{\specK}[1]{\ensuremath{\textcolor{blue}{#1}}}

\newcommand{\spec}[1]{\specK{\left\{{#1}\right\}}}

\newcommand{\dotcup}{\ensuremath{\mathaccent\cdot\cup}}

\newcommand{\pcmF}{\bullet}
\newcommand{\join}{\pcmF}

\newcommand{\own}{\mathsf{own}}
\newcommand{\nown}{\cancel{\own}}
\newcommand{\valid}{\mathsf{defined}}
\newcommand{\hflat}[1]{\ulcorner{#1}\urcorner}
\newcommand{\zag}{\lhd}
\newcommand{\zig}{\rhd}

\newcommand{\sfst}[1]{#1\backslash 1}
\newcommand{\ssnd}[1]{#1\backslash 2}
\newcommand{\sproj}[2]{#2\backslash{#1}}
\newcommand{\spair}[2]{[#1, #2]}
\usepackage{amsmath}
\makeatletter
\newcommand\ostep[2][]{\ext@arrow 0099{\longrightarrowfill@}{#1}{#2}}
\def\longrightarrowfill@{\arrowfill@{\ \ \ }\relbar\longrightarrow}
\makeatother

\makeatletter
\newcommand\osteps[2][]{\ext@arrow 0099{\longrightarrowfillstar@}{#1}{#2}}
\def\longrightarrowfillstar@{\arrowfill@{\ \ \ }\relbar{\longrightarrow^*}}
\makeatother

\makeatletter
\newcommand\mstep[2][]{\ext@arrow 0099{\longrightarrowfill@}{#1}{#2}}
\def\longrightarrowfill@{\arrowfill@{\ \ \ }\relbar\longrightarrow}
\makeatother

\makeatletter
\newcommand\msteps[2][]{\ext@arrow 0099{\longrightarrowfillstar@}{#1}{#2}}
\def\longrightarrowfillstar@{\arrowfill@{\ \ \ }\relbar{\longrightarrow^*}}
\makeatother

\newcommand{\stab}[1]{#1^\bullet}
\newcommand{\fstab}[2]{#1^{#2}}
\newcommand{\smorph}[2]{{#1}\hat{~}{#2}}

\newcommand{\rightarrowX}[1]{\overset{#1}{\rightarrow}}

\newcommand{\bwedge}{\,{\boldsymbol{\wedge}}\,}
\newcommand{\barrow}{\,{\boldsymbol{\rightarrow}}\,}
\newcommand{\bstar}{\,{\boldsymbol{*}}\,}

\newcommand{\dotof}[2]{{#1}{.}{#2}}
\newcommand{\Sof}[1]{\dotof{#1}{S}}
\newcommand{\Flatof}[2]{\dotof{#1}{\hflat{#2}}}
\newcommand{\TransSet}{\Delta}
\newcommand{\IntSet}{\TransSet_i}
\newcommand{\ExtSet}{\TransSet_e}
\newcommand{\transof}[1]{\dotof{#1}{\TransSet}}
\newcommand{\intof}[1]{\dotof{#1}{\IntSet}}
\newcommand{\extof}[1]{\dotof{#1}{\ExtSet}}

\setcopyright{none}
\renewcommand\footnotetextcopyrightpermission[1]{} 
\begin{document}

\title{Subjective Simulation as a Notion of Morphism for Composing Concurrent Resources}
\author{Aleksandar Nanevski}
\email{aleks.nanevski@imdea.org}
\affiliation{IMDEA Software Institute}
\author{Anindya Banerjee}
\email{anindya.banerjee@imdea.org}
\affiliation{IMDEA Software Institute}
\author{Germ\'an Andr\'es Delbianco}
\email{gad@irif.fr}
\affiliation{IRIF - Universit\'e Paris Diderot}
\begin{abstract}

Recent approaches to verifying programs in separation
logics for concurrency have used state transition systems (STSs) to
specify the atomic operations of programs.
A key challenge in the setting has been to compose such STSs into
larger ones, while enabling programs specified under one STS to
be lifted to a larger one, without reverification.
This paper develops a notion of morphism between two STSs which
permits such lifting. The morphisms are a constructive form of
simulation between the STSs, and lead to a general and concise proof
system. We illustrate the concept and its generality on several
disparate examples, including staged construction of a readers/writers
lock and its proof, and of proofs about quiescence when concurrent
programs are executed without external interference.

\end{abstract}

\maketitle
\thispagestyle{empty}


\newcommand{\internal}{\emph{internal}\xspace}
\newcommand{\external}{\emph{external}\xspace}
\newcommand{\SCSL}{\textsc{SCSL}\xspace}
\newcommand{\FCSL}{\textsc{FCSL}\xspace}
\newcommand{\lift}{\textsc{Lift}\xspace}

\section{Introduction}\label{sec:intro}
In many separation logics for shared-memory concurrent programs, a
formal description of a concurrent resource takes a form of a state
transition system
(STS)~\cite{DinsdaleYoung-al:ECOOP10,Nanevski-al:ESOP14,Jung-al:POPL15}.
The state space of an STS describes what holds of the resource's heap
and auxiliary state at all times during execution, while the
transitions specify the moves that programs operating over the
resource are allowed to make atomically. Thus, resources are part of
program specification: when verifying a program that operates over a
resource, one not only has to establish the program's pre- and
postcondition, but also show that the program respects the resource's
state space and transitions. In the sequel, we use ``resource'' and
``STS'' interchangeably.\footnote{Related works have also used names
  such as \emph{concurrent protocols}, \emph{distributed protocols},
  and \emph{concurroids} for similar concepts.}
%
%
%

One of the major challenges of the approach---which we address in this
paper---has been to design a formalism for composing resources
into new ones, which, moreover, allows the reuse of proofs carried out
for programs written for constituent resources, as follows. Once
resources are composed, it should be possible to \emph{lift} a program
that has been verified wrt.~one of the component STSs, and
automatically infer its correctness wrt.~the composition, without any
re-verification.
%
%

Consider the example of a concurrent resource in the
style of Concurrent Separation Logic (CSL)~\cite{OHearn:TCS07}. This
is a lock-protected shared heap satisfying a predicate, say $I$,
(aka.~resource invariant~\cite{Owicki-Gries:CACM76}) when no thread
holds the lock.
When the lock is acquired, the protected heap is transferred to the
exclusive ownership of the acquiring thread. While in exclusive
possession of the heap, the thread can modify the heap to temporarily
violate $I$, but has to re-establish $I$ before unlocking, when the
heap becomes shared again. 

CSL is coarse-grained, locking the whole data structure before
modification. Nevertheless, it already illustrates the need for
decomposition. A CSL-style resource performs two distinct
functionalities: locking and unlocking on the one hand, and
transferring heap ownership on the other. The two problems have
separate concerns and can appear individually in different contexts.
For example, transfer of heap ownership occurs when a concurrent
stack operation allocates a new node in a private state, and then
pushes it onto the shared stack, without actually locking the whole
structure. Similarly, locking and unlocking may be considered
independently of ownership transfer, or in settings where the
ownership discipline is more involved than in CSL. For example, in
readers-writers lock~\cite{courtois:ACM71,Bornat-al:POPL05}, when a
reader acquires the lock, the protected heap is not transferred to the
private ownership of that reader, but can be shared by all readers in
the system. Thus, the two different functionalities are best
formalized as individual STSs, which can then be composed into a
CSL-style lock, or used separately.


\newcommand{\Spin}{\mathsf{Spin}}
\newcommand{\Xfer}{\mathsf{Xfer}}

However, to recover the CSL-lock functionality by composition, one
must interconnect the states and transitions of the two
components, as they are not independent. For example, let $\Spin$ be a
resource implementing a spin lock. We will formally describe this
resource in Section~\ref{sec:overview}, but, as a first approximation,
one may envision an STS with two states and two non-idle transitions,
lock and unlock. Next, let $\Xfer$ be a resource implementing the
ownership transfer of a heap, under resource invariant $I$. Again as
an approximation, $\Xfer$'s states consist of a private and a shared
heap, and the transitions move a set of pointers circumscribed by $I$
between the two heaps. To reconstitute a CSL lock as a composition of
$\Spin$ and $\Xfer$, we have to ensure that whenever $\Spin$
transitions by taking the lock, $\Xfer$ is able to transfer
the shared heap into private ownership of the locking thread:
this heap must not already be privately owned. Dually, whenever $\Spin$
transitions to release the lock, then $\Xfer$ must ensure that there
exists a chunk of private heap that satisfies invariant $I$ and that can be
transferred into the shared state. During either of these transitions
by $\Spin$, $\Xfer$ should not be able to perform any other
manipulation of the heap, and vice versa.

Moreover, if we write a \emph{program} over $\Spin$, we should be able
to lift it to operate on states that lie in the composition of $\Spin$
and $\Xfer$. For example, a program for locking may be implemented as
a loop trying to take a lock, until it succeeds. This program respects
the transitions of $\Spin$, because either it stays idle if it fails
to take the lock, or it makes the lock transition of $\Spin$ in the
loop's last iteration. Once this program is verified wrt.~$\Spin$, we
should be able to \emph{lift} it to work over the composition of
$\Spin$ and $\Xfer$, without additional proof obligations. Whenever
the program would have taken a transition of $\Spin$, the lifting has
to take a transition in the composition, i.e., transform an $\Xfer$
part of the composed state by a specific, possibly non-idle, $\Xfer$
transition.



The customary mathematical structure for relating STSs are
\emph{simulations}~\cite{aba+lam:91}. However, most modern separation
logics for concurrency, while using STSs to formalize resources,
relate the resources, and tie them to program lifting, by notions
other than simulations (see Section~\ref{sec:related}). Examples include \emph{higher-order auxiliary
  code}~\cite{Jacobs-Piessens:POPL11, Svendsen-al:ESOP13,
  Svendsen-Birkedal:ESOP14}, \emph{atomicity
  tokens}~\cite{ArrozPincho-al:ECOOP14, Jung-al:POPL15}, and
\emph{protocol hooks}~\cite{ser+al:oopsla17}, among others.
In practice, the use of each of these concepts leaves one with a sense that
there is a simulation between underlying resources that is being
implicitly constructed; but the simulation is never 
made an explicit object of the formalism. 

In contrast, this paper advocates a form of simulation between STSs as
a key concept to relate resources and formalize program lifting.
%
%
If a resource $V$ is a sub-component of $W$, as in the above example
of $\Spin$ and CSL-style lock, then $W$ simulates $V$. Then, a program
$e$ operating over $V$ can easily be lifted to operate over $W$:
whenever $e$ takes a transition of $V$, the lifted program should take
a corresponding transition of $W$, which is guaranteed to exist
because of the simulation. The \textbf{fundamental contribution} of
this paper is this notion of simulation as a foundation for separation
logics for concurrency. 
Specifically, we develop a new logic
%
%
which reformulates previous work on Fine-grained
Concurrent Separation Logic (\FCSL)~\cite{LeyWild-Nanevski:POPL13,
  Nanevski-al:ESOP14}. The new logic, also called \FCSL, is designed
around simulations 
to achieve significant conceptual and formal simplicity compared to
the previous work on \FCSL, or the other related works listed
above. For example, we require only a \emph{single inference rule} to
reason about program lifting.


There are several hurdles to overcome in the design of \FCSL, leading
to the two main technical contributions of this paper.  First, we must
focus on a special kind of simulations, that are \emph{constructive}
in the sense of type theory. Whenever $V$ can take a transition, it
does not suffice merely to know that there \emph{exists} a transition
that $W$ can take as well; we need a witness for the existential. Only
then can we use our simulation as a \emph{morphism} on programs, that
is, a function that can modify a program over $V$ on-the-fly, into a
program over $W$. Our \textbf{first technical contribution} is to
identify the properties that make a simulation be a morphism, in the
above sense.
%

In more detail, the new \FCSL Hoare triples have the form of a
typing judgment
$e : \spec{P}\ A\ \spec{Q}@V$.
The judgment states that program $e$ returns a value of type $A$ (if it
terminates), $e$ respects the state space and transitions of $V$, and
has precondition $P$ and postcondition $Q$, assuming interference that
also respects the state space and transitions of $V$.
A morphism $f : V \rightarrow W$ is a structure that relates the
states of $V$ and $W$, and maps the transitions of $V$ to transitions
of $W$. The following single inference rule lifts program $e$ over $V$ to
program $\mathsf{morph}~f~e$ over $W$ by applying $f$ to $e$:
%
\begin{mathpar}
\inferrule*[right=\lift]{e : \spec{P}~A~\spec{Q}@V}
          {\mathsf{morph}~f~e : \spec{\smorph f P \bwedge I}\ A\ \spec{\smorph f Q \bwedge I}@W}
\end{mathpar}
%
Intuitively, the behavior of $\mathsf{morph}\ f\ e$ is to take the
transition $f(t)$ in $W$, whenever $e$ takes the transition $t$ in
$V$. And, $\smorph f P$ is the action of $f$ on predicates over state,
defined as $\smorph f P = \lambda s_w.\, \exists s_v.\, (s_v, s_w) \in
f \wedge P\ s_v$, where $s_v$ and $s_w$ are states from the state
spaces of $V$ and $W$, respectively.\footnote{In separation logic,
  logical connectives operate on state predicates. Here, we make a
  typographic distinction between predicate connectives (bold font),
  and propositional connectives (regular font). For example, $P
  \bwedge Q = \lambda s\ldot P\ s \wedge Q\ s$.}
%
$I$ is a predicate over states of $W$, which is ``preserved'' by $f$
in a sense that we formally define in Section~\ref{sec:formal}.



Soundness considerations of the above rule lead to our \textbf{second
  technical contribution}, which is novel structure on resource
transitions.
In previous work on \FCSL, a state of a resource distinguished between
\emph{self}-components (private to the specified thread), and
\emph{other}-components (private to the interfering threads). The
\emph{other}-component abstracted from the context of interfering
threads, making it unnecessary to reverify programs when the number of
interfering threads changed~\cite{LeyWild-Nanevski:POPL13}. This state
organization was named \emph{subjective}, because it gave each thread
its local (i.e., subjective) view of state ownership.
In contrast, this paper extends the subjective dichotomy to
transitions, and differentiates between \internal and \external
transitions. The \internal transitions of resource $V$ are those that a
program over $V$ can take. The \external transitions cannot be taken
by a program directly, but they delimit how $V$ can be combined with
other resources, and in particular, how a thread over the combined
resource can interfere with a thread over $V$. \emph{External} 
transitions thus abstract from the resource context in which $V$
appears, and serve as $V$'s interface.
%
%
%
%
A morphism $f : V \rightarrow W$ is a simulation that treats
\emph{self} components and \internal transitions differently from
\emph{other} components and \external transitions, as
follows. 
\begin{enumerate}
\item Every \internal transition $t$ of $V$ is matched by an \internal
  transition $f(t)$ of $W$, modifying \emph{self}-components, but
  preserving \emph{other}-components.
\item 
Every \external transition of $W$ is matched by \emph{one or
  more} transitions of $V$, of either kind, in succession, modifying
\emph{other}-components, but preserving \emph{self}-components.
\end{enumerate}
Requirement (1) ensures that $\mathsf{morph}~f~e$ lifts the atomic
steps of $e$ from $V$ to $W$.  Requirement (2) ensures that atomic
steps performed by interfering threads to $\mathsf{morph}~f~e$ over
$W$, can also be seen as atomic steps performed by interfering threads
to $e$ over $V$. Together, the requirements enable exploiting the
Hoare type of $e$ in the premiss of the \lift rule, and ensuring the
latter's soundness.

The notion of morphism has applications that go beyond resource
composition and lifting. For example, Section~\ref{sec:formal} shows
how to add a new property $I$ to the state space of a resource $V$, so
long as $I$ is inductive (i.e., preserved by $V$'s
transitions). Moreover, there is a \emph{generic} morphism from $V$ to
the restricted resource $V/I$.
%
Section~\ref{sec:param} illustrates how to use
morphisms in a generalized form of \emph{indexed morphism families},
to formalize \emph{quiescence}~\cite{Nanevski-al:ESOP14,
  sergey:oopsla16}. This is a situation when a resource $V$ is
installed in a private state of some program $e$. The children threads
of $e$ may compete for the new resource, but other threads cannot
interfere, because they cannot access $e$'s private state.
%

All our examples (including ones not discussed in the paper) and meta
theory have been mechanized in Coq, and the sources are available in
the supporting material.

\newcommand{\Shar}{\mathsf{Shar}}
\newcommand{\Priv}{\mathsf{Priv}}
\newcommand{\CSL}{\mathsf{CSL}}
\newcommand{\unlock}{\pi}

\section{Overview}\label{sec:overview}
We introduce \FCSL by developing CSL-style locks in a decomposed
manner.  The resource $\Spin$ formalizes locking over the spin lock
$r$. The resource $\Xfer$ formalizes ownership transfer of the
protected heap, enforcing that a resource invariant $I$ holds of the
heap when it is shared. The resource $\CSL$ composes $\Spin$ and
$\Xfer$, enforcing that: (1) when $\Spin$ locks, $\Xfer$ enables the
heap to be acquired by the locking thread, and (2) $\Spin$ unlocks
only after $\Xfer$ has been placed in a state whereby $I$ holds of the
heap. A morphism $f : \Spin \rightarrow \CSL$ can lift $\Spin$
programs for locking and unlocking to $\CSL$, thereby reusing the
programs' code and proof in $\Spin$.

\subsection{Resource $\Spin$ for locking and unlocking}\label{sec:spin}
Physically, a spin lock is a Boolean pointer $r$, which is locked if
$r$ is $\mathsf{true}$. Threads try to lock by executing
$\mathsf{CAS}(r, \mathsf{false}, \mathsf{true})$. The latter reads
from $r$, and, if $\mathsf{false}$, sets $r$ to $\mathsf{true}$,
returning $\mathsf{true}$ to indicate successful locking. We assume
that memory operations over a single pointer are atomic; thus, no
threads can modify $r$ between the reading and mutation by
$\mathsf{CAS}$. A thread that holds $r$, releases it by writing
$\mathsf{false}$ into it.
For verification, however, $\Spin$ cannot comprise only the boolean
states indicating whether $r$ is locked or not. It has to additionally
track which thread, if any, actually holds $r$, as such threads will
be allowed operations not allowed to others (e.g., unlocking). One way
to track lock ownership is by thread id's, but we do not do so
here. Instead, we endow $\Spin$ with a special form of
\emph{subjective} state (Section~\ref{sec:intro}), described
concretely below. As we shall see, subjective state will apply to all
our resources, with uses well beyond replacing thread
id's~\cite{LeyWild-Nanevski:POPL13,Nanevski-al:ESOP14}.

\paragraph{Subjective states.} 
We divide the state $s$ of $\Spin$ into three components $s =
(\thelock_s, \unlock, \thelock_o)$. Each thread over $\Spin$ has these
components in its name-space, but they may have different values in
different threads. For example, the \emph{self}-component $\thelock_s$
equals $\own$ in the thread that holds the lock, but $\nown$ in all
other threads. Dually, the \emph{other}-component $\thelock_o$ equals
$\own$ in a thread whose environment holds the lock, and $\nown$
otherwise. The lock is taken if exactly one of $\thelock_s$ and
$\thelock_o$ is $\own$. Importantly, each thread is allowed to modify
only its own $\thelock_s$ value, but not $\thelock_o$, and dually,
$\thelock_s$ of one thread cannot be changed by others. This way, the
division into \emph{self} and \emph{other} fields captures a form of
ownership.
On the other hand, the $\unlock$ component is under \emph{joint}
(i.e., shared) ownership. We introduce it with the view towards the
composition of $\Spin$ and $\Xfer$, and it is a Boolean indicating
that the invariant $I$ holds of the heap in $\Xfer$. This heap is not
part of $\Spin$, so $\unlock$ is essentially a proxy that will be
ascribed the explained meaning only after we compose $\Spin$ and
$\Xfer$. For now, it suffices to consider $\unlock$ as a field that a
thread wanting to unlock $r$ must set to $\mathsf{true}$, in
addition to having $\thelock_s = \own$.\footnote{It is customary in
  separation logic to refer to $\pi$ as a \emph{permission} to
  unlock. We refrain from doing so, as for us $\unlock$ is a
  necessary, but not sufficient condition for unlocking, as the thread
  must also set $\thelock_s = \own$.} In the sequel, we treat the
field names as projections, and write, for example, $\thelock_s(s)$
and $\thelock_s(s')$, when we want to extract the first component of
the states $s$ and $s'$, respectively.


The fields $\thelock_s$, $\thelock_o$, and $\unlock$ must be related
by some conditions, which we describe next.
First, 
%
%
we define the operation $\join$ on $O = \{\own, \nown\}$ as follows:
$x \join \nown = \nown \join x = x$ with $\own \join \own$
undefined. The operation is commutative, associative, with $\nown$ as
the unit element, hence it endows $O$ with the structure of a partial
commutative monoid
(PCM)~\cite{LeyWild-Nanevski:POPL13,DinsdaleYoung-al:POPL13,Nanevski-al:ESOP14,Jung-al:POPL15}. We
can now abbreviate $\thelock(s) = \thelock_s(s) \join \thelock_o(s)$
to capture the lock status; $r$ is taken iff $\thelock(s) = \own$.
%
%
Second, for each resource, we define its \emph{flattening}, which maps
the abstract state $s$ into a heap $\hflat{s}$, thereby declaring that
the values $\thelock_s$, $\thelock_o$ and $\pi$ are
\emph{auxiliary}~\cite{Lucas68,Owicki-Gries:CACM76}---they are
introduced for verification, but do not matter in execution, where
only $\hflat{s}$ matters.
%
%
%
Now we can define the state space of $\Spin$, which relates
$\thelock_s$, $\thelock_o$ and $\unlock$ as follows.
\[
\begin{array}{rcl}
S(s) & \eqdef & \valid~(\thelock(s)) \wedge r \neq \mathsf{null} \wedge 
    (\thelock(s) = \nown \rightarrow \unlock(s))\\
\hflat{s} & \eqdef & r \hmapsto (\thelock(s) \eqbool \own)
\end{array}
\]
The conjunct $\valid~(\thelock(s))$ encodes mutual exclusion: two
different threads cannot simultaneously hold the lock because if
$\thelock_s(s) = \thelock_o(s) = \own$, then $\thelock(s)$ would be
undefined. The conjunct $r \neq \mathsf{null}$ requires that $r$ is a
valid heap pointer. The last conjunct in $S(s)$ says that if the lock
is free, then, in the eventual composition with $\Xfer$, the protected
heap of $\Xfer$ satisfies the invariant $I$, thus encoding the main
property of CSL-style locking.  The definition of $\hflat{s}$ declares
that $\Spin$'s physical heap contains only the lock $r$, which is
locked if $\thelock(s) = \own$.
%

%

\paragraph{Transitions} A transition is a binary relation between
a pre-state $s$ and post-state $s'$, formalizing the atomic operations
of a resource. 
In the display below, we present the transitions of $\Spin$, where we
assume that both states $s$ and $s'$ satisfy $\Spin$'s $S$.
%
%
\[
\begin{array}{rcl}
\mathsf{lock\_tr}\ s\ s' & \eqdef & \thelock(s) = \nown \wedge \thelock_s (s') = \own \wedge \unlock(s')\\
\mathsf{unlock\_tr}\ s\ s' & \eqdef & \thelock_s(s) = \own \wedge \unlock(s) \wedge \thelock_s(s') = \nown\\
\mathsf{set\_tr}\ b\ s\ s' & \eqdef & \thelock_s(s) = \thelock_s(s') = \own \wedge \unlock(s') = b\\
\mathsf{id\_tr}\ P\ s\ s' & \eqdef & P\ s \wedge s' = s
\end{array}
\]
Transition $\mathsf{lock\_tr}$ describes a \emph{successful}
acquisition of the lock. It can be taken only if the lock is free
($\thelock(s)=\nown$), and in the post-state, the lock is held by the
acquiring thread ($\thelock_s(s') = \own$). By definition of $S$,
$\unlock$ must be set in $s$, and it remains so in $s'$.
On the other hand, $\mathsf{set\_tr}$ takes a boolean $b$ as an input,
and sets $\unlock$ to $b$. It can be performed only by a thread that
holds the lock ($\thelock_s(s) = \own$).
Similar explanation applies to $\mathsf{unlock\_tr}$ which describes
unlocking. Notice that transitions may modify $\thelock_s$ and
$\unlock$, but can only read $\thelock_o$, as the latter is owned by
other threads.
It is therefore always the case in a transition that $\thelock_o(s') =
\thelock_o(s)$, which we thus assume as default, and omit stating
explicitly.
%
%
The idle transition $\mathsf{id\_tr}$ is taken by a thread when it executes no state
changes, i.e., it stays idle. We parametrize $\mathsf{id\_tr}$ by a
predicate $P$, to describe what holds of the pre-state when the
transition is taken.  As we show promptly, this will be exploited when
defining the \emph{action} for locking, when the idle transition will
describe when the locking \emph{fails}. If $P$ is the always-true
predicate, we omit it.

%

\paragraph{Actions} 
Transitions describe the steps of a resource at the level of
specification, while \emph{actions} describe the atomic operations
\emph{at the level of programs}. Actions are composed out of one or
more transitions, and return a result that identifies the transition
taken by the action. Thus, an action is a relation between the output
result, the input state, and the output state.
For example, the action $\mathsf{trylock\_act}$ takes the transition
$\mathsf{lock\_tr}$ in the case of successful locking, and
$\mathsf{id\_tr}\ P$ otherwise. We use $P \eqdef \lambda s\ldot
\thelock(s) = \own$ to indicate that the locking fails only if the
lock were taken in $s$.
\[
\mathsf{trylock\_act}\ (b : bool)\ s\ s' \eqdef
\act{if}\ b\ \act{then}\ \mathsf{lock\_tr}\ s\ s'\ \act{else}\ \mathsf{id\_tr}\ (\lambda
s \ldot \thelock(s) = \own)\ s\ s'.
\]
While $\mathsf{trylock\_act}$ is defined over the whole state of
$\Spin$, including auxiliary values such as $\thelock(s)$, notice that
when the state is flattened to the pointer $r$, the action,
intuitively, behaves like $\act{CAS}(r, \mathsf{false},
\mathsf{true})$ discussed before.  We say that $\mathsf{trylock\_act}$
\emph{erases} to $\mathsf{CAS}$, or alternatively, that
$\mathsf{trylock\_act}$ annotates $\mathsf{CAS}$ with \emph{auxiliary
  code} for updating $\thelock_s$, $\thelock_o$ and $\unlock$. All our
actions erase to some memory operation that executes atomically on
hardware.


The action $\mathsf{unlock\_act}$ does not branch, but takes
the $\mathsf{unlock\_tr}$ transition, returning the result of unit
type. The action $\mathsf{unlock\_act}$ erases to the atomic operation
of writing $\mathsf{false}$ into $r$.
\[
\mathsf{unlock\_act}\ (x:unit)\ s\ s' \eqdef \mathsf{unlock\_tr}\ s\ s'
\]

We can now implement the programs for locking and unlocking
$r$.\footnote{The proofs of the type ascriptions are in our Coq
  files.}  The former loops executing $\mathsf{trylock\_act}$ until it
succeeds to acquire $r$, while the latter just invokes
$\mathsf{unlock\_act}$.
\[
\begin{array}[t]{l}
\mathsf{lock} : \begin{array}[t]{l}
          \spec{\lambda s\ldot \top}\ 
          \spec{\lambda s\ldot \thelock_s(s) = \own \wedge \unlock(s)}@\Spin = \hbox{}\\
          \end{array} \\
\quad \act{do}\ (\!\!\!\begin{array}[t]{l}
 b \leftarrow \mathsf{atomic}~\mathsf{trylock\_act};\\ 
 \act{if}\ b\ \act{then}\ \act{ret}\ ()\ \act{else}\ \mathsf{lock})
\end{array}
\end{array} 
\quad
\begin{array}[t]{l}
\mathsf{unlock} : \begin{array}[t]{l}
  \spec{\lambda s\ldot \thelock_s (s) = \own \wedge \unlock(s)} \\
  \spec{\lambda s\ldot \thelock_s (s) = \nown}@ \Spin = \hbox{}
  \end{array}\\
\quad \act{do}\ (\mathsf{atomic}~\mathsf{unlock\_act})
\end{array}
\]
The precondition of $\mathsf{lock}$ is $\top$, hence $\mathsf{lock}$
can be invoked in any state. The postcondition indicates that the lock
is acquired by the invoking thread, and $\unlock$ is set. This holds
because the program loops, until it manages to execute
$\mathsf{lock\_tr}$, which terminates with the lock acquired and
$\unlock$ set. 
%
The precondition of $\mathsf{unlock}$ requires the invoking thread to
hold the lock, and $\unlock$ to be set. Upon termination, the thread
does not have the lock anymore, as expected, but also notice that
$\unlock$ is undetermined. $\mathsf{Unlock\_tr}$ terminates with
$\unlock$ set, thus, immediately upon execution of $\mathsf{unlock}$,
we know that $\unlock$ will be set. However, our specifications state
only \emph{stable} properties of state, i.e., those that remain
invariant under interference of other threads over $\Spin$. In this
particular case, another thread may reset $\unlock$ after
$\mathsf{unlock}$ terminates, which is why $\unlock$ is undetermined
in $\mathsf{unlock}$'s postcondition. On the other hand, $\unlock$
holds stably in $\mathsf{lock}$'s postcondition because only the
thread holding the lock can reset $\unlock$.

%
%
%

\subsection{Resource $\Xfer$ for heap ownership transfer}\label{sec:xfer}

\newcommand{\here}{\nu}

A state $s$ of $\Xfer$ has the form $s = (\theheap_s, (\theheap_j,
\here), \theheap_o)$. The fields $\theheap_s$ and $\theheap_o$
describe the private heaps of the thread operating over $\Xfer$, and
the thread's environment, respectively. The field $\theheap_j$ is the
shared heap on which we consider the satisfaction of the resource
invariant $I$. Heaps form a PCM under the operation of disjoint union,
with $\mathsf{empty}$ as unit, just as was the case with the
\emph{self} and \emph{other} fields in  $\Spin$; we
abbreviate the total heap of $s$ as $\theheap(s) = \theheap_s(s) \join
\theheap_j(s)\join \theheap_o(s)$. The field $\here$ is a boolean
indicating the satisfaction of the invariant. The state space of $\Xfer$ is defined as follows.
\[
\begin{array}{rcl}
S(s) & = & \valid~(\theheap(s)) \wedge \act{if}\ \here(s)\ \act{then}\ I~\theheap_j(s)\ \act{else}\ \theheap_j(s) = \mathsf{empty}\\
\hflat{s} & = & \theheap(s)
\end{array}
\]
Specifically, if $\here$ is $\mathsf{true}$, then $I$ holds of
$\theheap_j$. Otherwise, the contents of $\theheap_j$ have been
transferred to $\theheap_s$ of some thread, and thus $\theheap_j$
equals $\mathsf{empty}$ heap.
%

The transitions of $\Xfer$ describe the exchange of heaps between
$\theheap_j$ and $\theheap_s$. We name them $\mathsf{close\_tr}$ and
$\mathsf{open\_tr}$, because they close and open the invariant $I$ for
violation, by moving a heap satisfying $I$ into and out of
$\theheap_j$.
\[
\begin{array}{rcl}
\mathsf{close\_tr}\ s\ s' & \eqdef & \exists h\ldot \theheap_s(s) = \theheap_s(s') \join h \wedge I\ h \wedge \neg \here(s) \wedge \theheap_j(s') = h \wedge \here(s')\\
\mathsf{open\_tr}\ s\ s' & \eqdef & \here(s) \wedge \theheap_s(s') = \theheap_s(s) \join \theheap_j(s) \wedge \theheap_j(s') = \mathsf{empty} \wedge \neg \here(s')\\
\end{array}
\]
$\mathsf{Close\_tr}$ moves the subheap $h$ of $\theheap_s(s)$ into
$\theheap_j(s')$. The moved heap $h$ must satisfy $I$, as otherwise,
$s'$ will not satisfy $S$.
%
The transition sets $\here(s')$ to indicate the satisfaction of $I$ in
$s'$. Symmetrically, $\mathsf{open\_tr}$ moves $\theheap_j(s)$ into
$\theheap_s(s')$, thereby leaving $\theheap_j(s') =
\mathsf{empty}$. We elide here the few additional $\Xfer$ transitions,
such as $\mathsf{id\_tr}\ P$ (defined identically as in $\Spin$), and
the transitions for mutating, allocating, and deallocating pointers in
$\theheap_s$, as they are not essential for our present goal of
explaining resource composition and morphisms.

\subsection{Composing $\Spin$ and $\Xfer$ into $\CSL$}\label{sec:csl}

The resource $\CSL$ combines the functionalities of $\Spin$ and
$\Xfer$, and admits morphisms from both. Specifically, the morphism
from $\Spin$ will allow us to automatically lift $\mathsf{lock}$ and
$\mathsf{unlock}$ to $\CSL$.

A state of $\CSL$ pairs up the states of $\Spin$ and $\Xfer$,
point-wise in the \emph{self}, \emph{joint} and \emph{other}
components. In other words,
$s = ((\thelock_s, \theheap_s), (\unlock, (\theheap_j, \here)),
(\thelock_o, \theheap_o))$. We write $\sfst s$ (resp.~$\ssnd s$) for
the first (resp.~second) point-wise projection of $s$. Thus,
$\sfst s = (\thelock_s, \unlock, \thelock_o)$ is a state of $\Spin$,
and $\ssnd s = (\theheap_s, (\theheap_j, \here), \theheap_o)$ is a
state of $\Xfer$.
We exclude some state pairings, however, as the following definitions
indicate:
\[
\begin{array}{rcl}
S(s) & = & \Sof{\Spin}(\sfst{s}) \wedge \Sof{\Xfer}(\ssnd{s}) \wedge \valid~\hflat{s} \wedge \unlock(s) = \here(s)\\
\hflat{s} & = & \Flatof{\Spin}{\sfst{s}} \join \Flatof{\Xfer}{\ssnd{s}}
\end{array}
\]
In particular, we require that: (1) The paired states have disjoint
heaps, i.e.~the lock $r$ from $\Spin$ does not occur as a pointer in
$\theheap(s)$ in $\Xfer$. This is imposed by the conjunct
$\valid~\hflat{s}$; (2) The booleans $\unlock$ and $\here$ from the
component STSs must be equal in the composition. This provides
$\unlock$ with the intended semantics from Section~\ref{sec:spin},
whereby it allows unlocking only if the protected heap satisfies
$I$. Indeed, when $\unlock(s) = \here(s) = \mathsf{true}$, then
$I\ \theheap_j(s)$ by definition of $\Sof{\Xfer}$, and $\Spin$ can
invoke $\mathsf{unlock\_tr}$. Dually, when $\unlock(s) = \here(s) =
\mathsf{false}$, then $\theheap_j(s) = \mathsf{empty}$, as the
protected heap is in private ownership of the locking thread, where
$I$ may be violated. Correspondingly, $\Spin$ cannot invoke
$\mathsf{unlock\_tr}$. However, the states where $\unlock(s) \neq
\here(s)$ are of no interest, and are ruled out by $S$.

Transitions of $\CSL$ combine the transitions of $\Spin$ and
$\Xfer$, as follows, omitting $\mathsf{id\_tr}$ for brevity:
\[
\begin{array}{rcl}
\mathsf{lock\_tr} & = & \dotof{\Spin}{\mathsf{lock\_tr}} * \dotof{\Xfer}{\mathsf{id\_tr}}\\
\mathsf{unlock\_tr} & = & \dotof{\Spin}{\mathsf{unlock\_tr}} * \dotof{\Xfer}{\mathsf{id\_tr}}\\
\mathsf{close\_tr} & = & \dotof{\Spin}{\mathsf{set\_tr}(\mathsf{true})} * \dotof{\Xfer}{\mathsf{close\_tr}}\\
\mathsf{open\_tr} & = & \dotof{\Spin}{\mathsf{set\_tr}(\mathsf{false})} * \dotof{\Xfer}{\mathsf{open\_tr}}
\end{array}
\]
We formally define the operation $t_1 * t_2$ of \emph{coupling} of
transitions in Section~\ref{sec:formal}, but for now it suffices to
say that $t_1 * t_2$ \emph{simultaneously} takes $t_1$ over $\sfst{s}$
(a state of $\Spin$), and $t_2$ over $\ssnd{s}$ (a state of $\Xfer$).
Thus, $\mathsf{lock\_tr}$ performs the lock transition of $\Spin$,
while remaining idle on $\Xfer$, and similarly for
$\mathsf{unlock\_tr}$. On the other hand, $\mathsf{open\_tr}$ (and
$\mathsf{close\_tr}$ is similar) executes
$\dotof{\Xfer}{\mathsf{open\_tr}}$ to transfer the shared heap to
private ownership, resetting $\here(s)$ in the process.
$\dotof{\Spin}{\mathsf{set\_tr}(\mathsf{false})}$ has to be
simultaneously executed, in order to maintain $\unlock(s) = \here(s)$.

\subsection{Morphisms}\label{sec:morph}
We next construct the morphism $f : \Spin \rightarrow \CSL$ that will
allow us to lift $\mathsf{lock}$ and $\mathsf{unlock}$
(Section~\ref{sec:spin}) from $\Spin$ to $\CSL$, thereby reusing their
$\Spin$ implementation and proof. The morphism consists of two parts:
a relation on the states of $\Spin$ and $\CSL$, and a function mapping
the transitions of $\Spin$ to those of $\CSL$.
%
Given state $s$ of $\Spin$ and $s'$ of $\CSL$, the state-relation part
of $f$ is:
\[
(s, s') \in f \eqdef s = \sfst{s'},
\]
using that a $\CSL$-state is a pair of a $\Spin$ and $\Xfer$ state. 
%
%
The transition-map part of $f$ is defined as: 
\[
\begin{array}{rcl}
f (\dotof{\Spin}{\mathsf{lock\_tr}}) & \eqdef & \dotof{\CSL}{\mathsf{lock\_tr}}\\
f (\dotof{\Spin}{\mathsf{unlock\_tr}}) & \eqdef & \dotof{\CSL}{\mathsf{unlock\_tr}}\\
f (\dotof{\Spin}{\mathsf{id\_tr}\ P}) & \eqdef & \dotof{\CSL}{\mathsf{id\_tr}~(\lambda s\ldot P\ \sfst{s})}\\
f (\dotof{\Spin}{\mathsf{set\_tr}\ b}) & \eqdef & \mathsf{undefined}
\end{array}
\]
The key role of $f$ is to establish a simulation between $\Spin$ and
$\CSL$, i.e., whenever $\Spin$ takes a transition $t$, $\CSL$ can take
a transition $f(t)$, with the input states of $t$ and $f(t)$ being
related by the state-relation of $f$, and similarly for the output
states.
When
$t \in \{\dotof{\Spin}{\mathsf{lock\_tr}},
\dotof{\Spin}{\mathsf{unlock\_tr}}, \dotof{\Spin}{\mathsf{id\_tr}}\
P\}$, it is easy to see that this property holds. For example, if
$t = \dotof{\Spin}{\mathsf{lock\_tr}}$, then
$f(t) = \dotof{\CSL}{\mathsf{lock\_tr}} =
\dotof{\Spin}{\mathsf{lock\_tr}} * \dotof{\Xfer}{\mathsf{id\_tr}}$.
When $t$ can be taken in $\Spin$, clearly $f(t)$ can be taken in
$\CSL$, since $\dotof{\Xfer}{\mathsf{id\_tr}}$ does not impose any
additional constrains.

Importantly, it is \emph{not possible} to make this property hold for
$t = \dotof{\Spin}{\mathsf{set\_tr}}$.  We could consider defining $f$
on $t$ as, e.g.,
$f(\dotof{\Spin}{\mathsf{set\_tr}\ (\mathsf{true})}) =
\dotof{\CSL}{\mathsf{close\_tr}} =
\dotof{\Spin}{\mathsf{set\_tr}(\mathsf{true})} *
\dotof{\Xfer}{\mathsf{close\_tr}}$, but such a definition does not
give a simulation. Namely, it is \emph{not} the case that when
$\dotof{\Spin}{\mathsf{set\_tr}(true)}$, then
$\dotof{\Xfer}{\mathsf{close\_tr}}$ can follow, as the latter requires
a further condition that there exist subheap $h$ of $\theheap_s(s)$
such that $I\ h$ holds. The existence of $h$ is not guaranteed by
$\dotof{\Spin}{\mathsf{set\_tr}(\mathsf{true})}$.

This motivates our \emph{division} of transitions into \internal and
\external, whereby morphisms are defined only on the \internal
ones. For $\Spin$, the \internal transitions are
$\dotof{\Spin}{\mathsf{lock\_tr}}$,
$\dotof{\Spin}{\mathsf{unlock\_tr}}$ and $\Spin.\mathsf{id\_tr}$, and
the \external transition is $\dotof{\Spin}{\mathsf{set\_tr}}$, on
which $f$ remains undefined.
Intuitively, 
%
%
\external transitions are ``incomplete'' operations, to be
``completed'' by the outside world, to which the \external transitions
are an interface. For example, $\dotof{\Spin}{\mathsf{set\_tr}}$ is
\external, because the very role of $\unlock$, which this
transition manipulates, is to tie $\Spin$ to another resource, in this
case $\Xfer$.
In the case of $\Xfer$, we similarly classify $\mathsf{close\_tr}$ and
$\mathsf{open\_tr}$ as \external, as they too are incomplete, but for
a somewhat different reason. Namely, an action involving these
transitions cannot be ascribed a stable Hoare triple in and of itself.
Indeed, a program trying to perform $\dotof{\Xfer}{\mathsf{open\_tr}}$
cannot rely that $\here(s)$ holds---and thus that there is a heap in
the shared state to be moved---as another simultaneous thread may
acquire the heap and reset $\here(s)$. This is avoided in
$\dotof{\CSL}{\mathsf{open\_tr}}$, which couples
$\dotof{\Xfer}{\mathsf{open\_tr}}$ with
$\dotof{\Spin}{\mathsf{set\_tr}(\mathsf{false})}$, and can thus be
executed only by a thread holding the lock. Hence, in $\CSL$,
$\mathsf{open\_tr}$ and similarly $\mathsf{close\_tr}$, are \internal.
\footnote{It is possible to make $\dotof{\Xfer}{\mathsf{open\_tr}}$
  stable, and thus internal, by introducing an additional field of
  type $O$ that tracks if a thread can execute the transition, and an
  additional external transition to manipulate the extra field. For
  simplicity, we do not explore such design here, but it is not
  precluded by the system.}
%

Since we want morphisms to act on programs such as $\mathsf{lock}$ and
$\mathsf{unlock}$ in Section~\ref{sec:spin}, the actions that a
program takes must be composed of \internal transitions only. For
example, programs $\mathsf{lock}$ and $\mathsf{unlock}$ use actions $\mathsf{trylock\_act}$ and
$\mathsf{unlock\_act}$, which are themselves defined in terms of
$\Spin$ transitions $\mathsf{lock\_tr}$, $\mathsf{unlock\_tr}$ and
$\mathsf{id\_tr}$, but not $\mathsf{set\_tr}$.
%
%
We can thus lift $\mathsf{lock}$ and $\mathsf{unlock}$ to $\CSL$, by
applying the $\lift$ rule with morphisms $f$ and
$I \eqdef \lambda s\ldot \theheap_s(s) = h$.
%
\[
\begin{array}{rcl}
\mathsf{lock'} & : & \specK{[h]\ldot{}}\spec{\lambda s\ldot \theheap_s(s) = h}\ \spec{\lambda s\ldot \thelock_s(s) = \own \wedge \here(s) \wedge \theheap_s(s) = h}@\CSL = \hbox{}\\
&& \quad  \mathsf{do}~(\mathsf{morph}\ f\ \mathsf{lock})\\
\mathsf{unlock'} & : & \specK{[h]\ldot{}}\spec{\lambda s\ldot \thelock_s(s) = \own \wedge \here(s) \wedge \theheap_s(s)=h}\ \spec{\lambda s\ldot \thelock_s(s) = \nown \wedge \theheap_s(s)=h}@\CSL = \hbox{}\\
&& \quad \mathsf{do}~(\mathsf{morph}\ f\ \mathsf{unlock})
\end{array}
\]
The operational intuition behind $\mathsf{lock'}$ (and
$\mathsf{unlock'}$ is similar) is that it executes $\mathsf{lock}$,
modifying $\mathsf{lock}$'s transitions by $f$. Program
$\mathsf{lock}$ loops executing $\dotof{\Spin}{\mathsf{id\_tr}}$,
until it finally executes $\Spin.\mathsf{lock\_tr}$. Accordingly,
$\mathsf{lock'}$ will keep executing $\dotof{\CSL}{\mathsf{id\_tr}}$
until it finally executes $\dotof{\CSL}{\mathsf{lock\_tr}}$, the
latter merely extending $\dotof{\Spin}{\mathsf{lock\_tr}}$ with
$\dotof{\Xfer}{\mathsf{id\_tr}}$. Thus, the specification of
$\mathsf{lock'}$ is similar to that of $\mathsf{lock}$ in that it
describes the modification to $\thelock_s$, but here it also states
that the private heap $\theheap_s(s)$ is unchanged from the
precondition to the postcondition, as in both, it equals the bound
variable $h$. The latter could not have been specified for
$\mathsf{lock}$, because the field $\theheap_s$ is not part of
$\Spin$, but is added by $\Xfer$.  In $\mathsf{lock'}$ we use
$\here(s)$ instead of $\unlock(s)$, as the two are equal by the
definition of $\CSL$'s state space.
In $\CSL$ we can further ascribe stable specification to
$\mathsf{close\_tr}$ and $\mathsf{open\_tr}$, since these are now
\internal transitions.
\[
\begin{array}{rcl}
\mathsf{close} & : & \specK{[h_1]\ldot{}}\!\!\!\begin{array}[t]{l}
  \spec{\lambda s\ldot \exists h_2\ldot \thelock_s(s) = \own \wedge \neg\here(s) \wedge \theheap_s(s) = h_1 \join h_2 \wedge I\ h_2}\\
  \spec{\lambda s\ldot \thelock_s(s) = \own \wedge \here(s) \wedge \theheap_s(s) = h_1}@\CSL = \hbox{}
  \end{array}\\
&& \mathsf{do}\ (\mathsf{atomic}~(\lambda x:unit\ldot \mathsf{close\_tr}))\\
\mathsf{open} & : & \specK{[h_1]\ldot{}}\!\!\!\begin{array}[t]{l}\spec{\lambda s\ldot \thelock_s(s)=\own \wedge \here(s) \wedge \theheap_s(s) = h_1}\\ 
                      \spec{\lambda s\ldot \exists h_2\ldot \thelock_s(s)=\own \wedge \neg\here(s) \wedge \theheap_s(s) = h_1 \join h_2 \wedge I\ h_2}@\CSL = \hbox{}\\
    \end{array}\\
&& \mathsf{do}\ (\mathsf{atomic}~(\lambda x:unit\ldot\mathsf{open\_tr}))
\end{array}
\]
We can then sequentially compose $\mathsf{lock'}; \mathsf{open}$ and
$\mathsf{close}; \mathsf{unlock'}$, to obtain programs that combine
lock operations with ownership transfer.

\subsection{Dividing $\Xfer$ into $\Shar$ and $\Priv$}\label{sec:shar}
It is very useful to further subdivide $\Xfer$ into two components
$\Shar$ and $\Priv$, which separately deal with shared heaps and
private heaps, respectively, and then inject each by means of a
morphism into $\Xfer$. 
%
%
%
$\Shar$ contains the fields $\theheap_j$ and $\here$, while $\Priv$
contains $\theheap_s$ and $\theheap_o$. Both have their own copies of
$\mathsf{give\_tr}$ and $\mathsf{trans\_tr}$ transitions which are
parametrized by the heap $h$. In the case of $\Shar$ (resp. $\Priv$),
these transitions describe how $h$ can be taken out of $\theheap_j$
(resp.~$\theheap_s$) or into it, but do not specify from which resource $h$
is received, or to which resource it is given away. Clearly, because they describe
interaction with the unspecified outside world, these transitions must
be \external.
\[
\begin{array}{rcl}
\dotof{\Shar}{\mathsf{take\_tr}}\ h\ s\ s' & \eqdef & I\ h \wedge \neg\here(s) \wedge \theheap_j(s') = h \wedge \here(s')\\
\dotof{\Shar}{\mathsf{give\_tr}}\ h\ s\ s' & \eqdef & h = \theheap_j(s) \wedge \here(s) \wedge \theheap_j(s') = \mathsf{empty} \wedge \neg\here(s')\\
\dotof{\Priv}{\mathsf{take\_tr}}\ h\ s\ s' & \eqdef & \theheap_s(s') = h \join \theheap_s(s)\\
\dotof{\Priv}{\mathsf{give\_tr}}\ h\ s\ s' & \eqdef & \theheap_s(s) = h \join \theheap_s(s')
\end{array}
\]
Dividing the functionality of $\Xfer$ will allow us to transfer the
shared heap $\theheap_j$ of $\Shar$ to some resource other than
$\Priv$. We will exploit this subdivision in Section~\ref{sec:rwlock}
on readers/writers, to facilitate reuse when formalizing different
heap ownership modes (i.e., heap owned by a writer vs. heap owned by
readers).

\makeatletter\def\@acmdefinitionindent{0pt}\makeatother
\makeatletter\def\@acmplainindent{0pt}\makeatother

\section{Formal structures}\label{sec:formal}

\subsection{Definitions}\label{sec:defs}

\begin{definition}[State-type and state]
A \emph{state-type} is a pair $(U, T)$ of a PCM $U$ and a type $T$. A
\emph{state} of state-type $(U, T)$ is a triple $s = (a_s, a_j, a_o)$
of type $U \times T \times U$. We use the labels as projections out of
$s$. The projections $a_s(s)$ and $a_o(s)$ of type $U$ are
called \emph{self} and \emph{other} component, respectively. The
projection $a_j(s)$ of type $T$ is called
\emph{joint} component. The \emph{self} component holds the values
that are private to the specified thread, and cannot be changed by
other threads. Dually, \emph{other} component holds the values that
are private to the environment of the specified thread, and cannot be
changed by the specified thread. The \emph{joint} component holds the
value that can be changed by every thread. 
\end{definition}

In a specific resource, we name the components with a
resource-specific name, but use $a_s$, $a_j$, $a_o$ when we discuss
resources in general. The $a_s(s)$ and $a_o(s)$ components of a state
$s$ present the local view of a thread that operates on $s$. Different
threads operating simultaneously on the same resource may have
different values for the $a_s$ and $a_o$ components of their states,
depending on the operations that they have completed. For example, in
Section~\ref{sec:overview}, a thread that acquired the lock will have
$a_s(s) = \thelock_s(s) = \own$, whereas a thread not holding the lock
will have $a_s(s) = \thelock_s(s) = \nown$. If these threads execute
at the same time, we further know that in the first thread $a_o(s)
= \thelock_o(s) = \nown$ and in the second, $a_o(s) = \thelock_o(s)
= \own$. 
In general, given any thread and a state $s$, the view of the whole
concurrent environment (i.e. all of the threads concurrent to the
considered thread), can be obtained by \emph{transposition} of $s$, as
per the following definition.

\begin{definition}[State transposition]
Given a state $s = (a_s, a_j, a_o)$, the \emph{transposition} of $s$ is
the state $s^\top = (a_o, a_j, a_s)$. 
\end{definition}

As customary in separation logic, a common operation in \FCSL is that
of \emph{framing}, i.e., adding values to state components. In \FCSL,
we consider framing of both of the PCM-valued components.


\begin{definition}[Two notions of framing]
Let $p \in U$ and $s$ be a state of state-type $(U, T)$. The
\emph{self-framing of $s$ with $p$} is the state $s \zag p = (a_s(s)
\join p, a_j(s), a_o(s))$. Dually, \emph{other-framing of $s$ with $p$} is
$s \zig p = (a_s(s), a_j(s), p \join a_o(s))$.
%
\end{definition}

A predicate is global if it is independent of the framing direction.
\begin{definition}[Globality]
Predicate $P$ over states of state-type $(U, T)$ is \emph{global}
if $P(s \zag p) \leftrightarrow P(s \zig p)$.
\end{definition}

Using again the notation from Section~\ref{sec:overview}, an example
of a global predicate is $P(s)\,{\eqdef}\,\thelock(s) = \own$. By
constraining the combined value $\thelock(s)
= \thelock_s(s) \join \thelock_o(s)$, $P$ says that the lock is taken,
but elides saying by whom. This is a general property; a global
predicate $P$ depends only on the combination $a_s(s) \join a_o(s)$,
but not on the individual values of $a_s(s)$ and $a_o(s)$. Indeed, by
definition, if $P$ is global, then $P (a_s, a_j, a_o) \leftrightarrow
P (a_s \join a_o, a_j, 1_U) \leftrightarrow P(1_U, a_j, a_s \join
a_o)$, where $1_U$ is the unit of the PCM $U$.
Thus, while $a_s$ and $a_o$ capture the effect on the resource by the
specified thread and by the concurrent environment, respectively, a
global predicate captures the total effect of all the threads,
ignoring which thread did exactly what.

Next, we define the properties of a resource state space. For example,
these will are satisfied by state spaces of $\Spin$, $\Xfer$ and
$\CSL$ from Section~\ref{sec:overview}.

\begin{definition}[State space]\label{def:coh}
\emph{State space} $S$ of state-type $(U, T)$ is a predicate over
states (equivalently, set of states) of state type $(U, T)$, that
satisfies the following properties: 
\begin{enumerate}
\item\label{ss:valid} \emph{(validity}) if $S(s)$ then $\valid\ (a_s(s) \join a_o(s))$
\item\label{ss:global} $S$ is global
\end{enumerate}
\end{definition}

Condition~(\ref{ss:valid}) in Definition~\ref{def:coh} captures that
we are only interested in states where the current thread and its
concurrent environment have jointly performed a valid effect over the
resource. For example, on Section~\ref{sec:overview}, this condition
imposes that we cannot have $\thelock_s(s) = \thelock_o(s) = \own$,
i.e., the lock cannot be simultaneously held by a thread and by its
environment. The globality condition~(\ref{ss:global}) closes up the
state-space under local views of simultaneous threads.  If two states
$s_1$ and $s_2$ are such that $a_j(s_1) = a_j(s_2)$ and
$a_s(s_1) \join a_o(s_1) = a_s(s_2) \join a_o(s_2)$, then $s_1$ and
$s_2$ represent the same moment in time of the resource, but from the
point of view of two different concurrent threads. $S$ being global
means that $S$ contains either both or neither of $s_1$ and
$s_2$. 



\begin{definition}[Flattening]\label{def:flat}
Let $S$ be a state space of state-type $(U, T)$. \emph{Flattening}
$\hflat{-} : S \rightarrow \mathsf{heap}$ is a function satisfying the
following properties.
\begin{enumerate}
\item\label{fl:valid} if $S(s)$ then $\valid~\hflat{s}$
\item\label{fl:zagzig} $\hflat{s\zag p} = \hflat{s \zig p}$
\end{enumerate}
When we want to emphasize the state space $S$, we write $S.\hflat{s}$
instead of $\hflat{s}$.
\end{definition}

Similarly to Definition~\ref{def:coh}, condition~(\ref{fl:valid})
captures that we only track resources whose flattened heap is valid,
i.e., it does not contain the $\mathsf{null}$ pointer, or duplicate
pointers. Condition~(\ref{fl:zagzig}) is similar to globality of $S$,
and says that flattening is independent of thread-local views.


\begin{definition}[State product]\label{def:stprod}
Let $s_i$ be states of state-types $(U_i, T_i)$, $i=1,2$. The product state
$\spair{s_1}{s_2}$ defined as
\[\spair{s_1}{s_2} \eqdef ((a_s(s_1), a_s(s_2)), (a_j(s_1),
a_j(s_2)), (a_o(s_1), a_o(s_2))\] 
is of state-type $(U_1 \times U_2, T_1 \times T_2)$, where $U_1 \times
U_2$ is a PCM with join and unit defined pointwise.
Symmetrically, given a state $s$ of state-type $(U_1 \times U_2,
T_1 \times T_2)$, the state $\sproj{i}{s}$ defined as $(\pi_i(a_s(s)),
\pi_i(a_j(s)), \pi_i(a_o(s))$ is of state-type $(U_i, T_i)$, $i=1,2$.
The usual beta and eta laws for products hold, i.e.:
$\sproj{i}{\spair{s_1}{s_2}} = s_i$ and $s = \spair{\sfst s}{\ssnd
s}$.
\end{definition}

\begin{definition}[State space product]\label{def:products}
Let $S_i$ be a state space of state-type $(U_i, T_i)$, $i=1,2$. Then
the following define a valid state space and flattening over the
product states:
\[
\begin{array}{rcl}
(S_1\times S_2)~s & \eqdef & S_1(\sfst{s}) \wedge S_2(\ssnd{s}) \wedge \valid~\hflat{s}\\
\hflat{s} & \eqdef & \Flatof{S_1}{\sfst{s}} \join \Flatof{S_2}{\ssnd{s}}
\end{array}
\]
The conjunct $\valid~\hflat{s}$ imposes that the
flattened heaps of component states are disjoint, in order to satisfy
the requirement of Definition~\ref{def:flat}.(\ref{fl:valid}).
\end{definition}

\begin{definition}[Transition]\label{def:trans}
Let $S$ be a state space of state-type $(U, T)$. \emph{Transition} $t$
over $S$ is a binary relation on states, satisfying the following
properties.
\begin{enumerate}
\item\label{def:func} \emph{(functionality)} if $t\ s\ s'_1$ and $t\ s\ s'_2$ then $s'_1 = s'_2$.
\item \emph{(other-fixity)} if $t\ s\ s'$, then $a_o(s) = a_o(s')$
\item \emph{(locality)} if $t\ (s \zig p)\ s'$ then there exists $s''$
  such that $s' = s'' \zig p$ and $t\ (s\zag p)\ (s''\zag p)$
\item\label{coh:preserve} \emph{($S$-preservation)} if $t\ s\ s'$ and $S(s)$ then $S(s')$
\end{enumerate}
When we want to emphasize the state space $S$ wrt.~which the
transition is defined, we write $S.t$ instead of $t$, and refer to $t$
as an $S$-transition.
We say that a state $s$ is \emph{safe} for a transition $t$, if there
exists $s'$ such that $t\ s\ s'$.
\end{definition}

Functionality requires that transitions are partial functions: the
output state of a transition may be undefined on some input state, but
if defined, it is unique.
%
%
Thus, transitions are deterministic operations. This includes
allocation, which separation logics often model
non-deterministically. In our Coq files, we implement a simple
concurrent allocator as a resource which keeps a free list, abstract
from the clients. The allocator deterministically models allocation
and deallocation by interacting with clients via transitions that
transfer the head pointer of the free list back and forth, much like
$\Xfer$ resource in Section~\ref{sec:overview} transferred a heap
between private and joint state.

Other-fixity 
captures that transitions
cannot change the other-view $a_o$ of a thread, which are read-only,
as already illustrated in Section~\ref{sec:overview}.

Locality 
is a form of frame property
from Abstract Separation Logic (ASL)~\cite{Calcagno-al:LICS07}. Let $s
= (a_s, a_j, a_o)$, and $s' = (a'_s, a'_j, a'_o)$, and assume that
$t~(s \zig p)~s'$. Ignoring \emph{joint} and \emph{other} components
for a moment, the assumption says that executing $t$ in a state with
the \emph{self} component $a_s$ results in a state with
the \emph{self} component $a'_s$. The locality property says that if
we increase the input \emph{self}-component to $a_s \join p$, then
the result and the increment are preserved; that is, the
output \emph{self} component is $a'_s \join p$. The specific of \FCSL,
compared to ASL, or other separation logics, is that the assumption
$t~(s \zig p)~s'$ requires the frame $p$ to be available in
the \emph{other} component of the input state. In this sense, locality
is a property stating an invariance of transitions under a
realignment of local views of threads, whereby we take a portion $p$
of the ``effect'' ascribed to an environment thread, and assign $p$ to
the specified thread.


Finally, the $S$-preservation property states that transitions
preserve the state space. We have tacitly assumed this property in the
examples in Section~\ref{sec:overview}.

\begin{definition}[Transition coupling]\label{def:coupling}
Let $t_i$ be an $S_i$-transition, $i=1,2$. Then \emph{coupling of
  $t_1$ and $t_2$} is the $(S_1\times S_2)$-transition $t_1 * t_2$,
defined as:
\[
(t_1 * t_2)\ s\ s' \eqdef t_1\ (\sfst{s})\ (\sfst{s'}) \wedge
t_2\ (\ssnd{s})\ (\ssnd{s'}) \wedge \valid\ \hflat{s'}
\]
\end{definition}
The coupled transition $t_1 * t_2$ executes $t_1$ and $t_2$
simultaneously, each on its respective portion of the input state.  By
the properties of $S_1 \times S_2$, we can assume that the input state
$s$ will have a valid flattening, i.e., that the heaps
$\hflat{\sfst{s}}$ and $\hflat{\ssnd{s}}$ are disjoint. However, when
$t_1$ and $t_2$ transition \emph{individually}, they might produce
respective ending states that share a common pointer (e.g., $t_1$ and
$t_2$ may receive the same pointer from the allocator). The conjunct
$\valid~\hflat{s'}$ prevents the coupled transition from ever
synchronizing $t_1$ and $t_2$ in such a way.


\begin{definition}[Internal transition]\label{def:internal}
An $S$-transition $t$ is \emph{internal} if it preserves the heap domain
of its input and output state; that is, whenever $t\ s\ s'$ then
$\hflat{s}$ and $\hflat{s'}$ contain the same pointers.
\end{definition}

Internal transitions are important because, intuitively, the set of
their safe states is not affected by coupling with other internal
transitions. More formally, if $s_1$, $s_2$ are safe for (internal)
$t_1$, $t_2$, respectively, and $\hflat{s_1}$ is disjoint from
$\hflat{s_2}$, then by Definition~\ref{def:internal},
$\spair{s_1}{s_2}$ is safe for $t_1 * t_2$.
%
%
We build atomic actions of programs out of internal transitions
only. Thus, the safety of a program whose atomic actions utilize the
internal transition $t_1$ will not be affected if $t_1$ is coupled
with an internal action $t_2$ over a disjoint state space. This
property is hence key for soundly lifting a program over one resource,
say $\Spin$, to a combined resource, say $\CSL$, which couples the
transitions of $\Spin$ with those of $\Xfer$.

{\emph{External}} transitions are not required to preserve heap
domains. External transitions describe interaction with other
resources, and enlarging or shrinking a resource's heap is a form of
interaction. For example, the transitions $\mathsf{take\_tr}$ and
$\mathsf{give\_tr}$ from Section~\ref{sec:shar}, acquire a new heap,
or give away a part of the existing heap, respectively.  External
transitions cannot be used to build actions \emph{directly}, but
external transitions of different resources can be coupled into an
internal transition of a combined resource, and then used in
actions. For example, coupling $\dotof{\Shar}{\mathsf{give\_tr}}~h$
and $\dotof{\Priv}{\mathsf{take\_tr}}~h$ in Section~\ref{sec:shar},
produces an effect of moving the heap $h$ from $\Shar$ to $\Priv$. But
in the combination $\Xfer$ of $\Shar$ and $\Priv$, this move is an
internal effect overall, essentially corresponding to the internal
transition $\dotof{\Xfer}{\mathsf{open\_tr}}$. 

\begin{definition}[Resource]\label{def:resource}
  A \emph{resource} (or \emph{STS}) is a tuple
  $V = (U, T, S, \IntSet, \ExtSet)$, where $S$ is a state space of
  state-type $(U, T)$, and $\IntSet$ and $\ExtSet$ are sets of
  internal and external $S$-transitions, respectively. We let
  $\TransSet = \IntSet \cup \ExtSet$ denote the set of all
  transitions. When $V$'s components are not explicitly named, we
  refer to them using the dot-notation. That is, $\dotof{V}{U}$ is
  $V$'s PCM, $\dotof{V}{T}$ is $V$'s type, etc. A state $s$ is a
  $V$-state, if it is of state-type $(\dotof{V}{U}, \dotof{V}{T})$.
\end{definition}

\begin{definition}[Inductivity]\label{def:inductive}
Let $V$ be a resource, and $I$ a predicate over $V$-states. We say
that $I$ is an inductive invariant for $V$, or $V$-inductive for
short, if it is preserved by the internal transitions of $V$; that
is: 
%
\begin{itemize}
\item for every $t \in \intof{V}$, if $t\ s\ s'$ and $I\ s$ then $I\ s'$.
\end{itemize}
\end{definition}

\begin{definition}[Other-stepping]\label{def:otherstep}
Let $V$ be a resource and $s$, $s'$ be $V$-states. We say that $s$
\emph{other}-steps by $V$ to $s'$, written $s \ostep[V]{} s'$, if
there exists a transition $t \in \transof{V}$ (thus, either internal
or external) such that $t\ s^\top\ s'^\top$. We write $\osteps[V]{}$
for reflexive-transitive closure of $\ostep[V]{}$.
\end{definition}

Because Definition~\ref{def:otherstep} uses transpositions of $s$ and
$s'$, the relation $s \osteps[V]{} s'$ expresses, from the point of
view of the specified thread, that $s$ can be modified into $s'$ by
the actions of the interfering threads. Other-stepping admits all
transitions in $\transof{V}$, not only the internal ones. We include
the external transitions to account for the possibility that a
resource can be modified by interfering programs that operate not over
$V$, but over some extension of $V$. For example, a heap in $\Priv$
may be augmented with another heap $h$ acquired from $\Shar$, once
$\Priv$ and $\Shar$ are combined into $\Xfer$.
%

\begin{definition}[Stability]\label{def:stable}
Let $V$ be a resource. Predicate $P$ over $V$-states is
\emph{stable in state $s$} if whenever $s\osteps[V]{}s'$, then
$P\ s'$. $P$ is \emph{stable} if it is stable in state $s$, for every
$s$ for which $P\ s$. 
%
Given $P$, we define its \emph{stabilization} $\stab{P}$ as 
$
\stab{P}\ s \eqdef \forall s'\ldot s \osteps[V]{} s' \rightarrow P\ s'.
$
It is easy to see that $\stab{P}$ is stable, and that $P$ is stable
iff $\forall s\ldot P~s \rightarrow \stab{P}~s$.
\end{definition}

For example, the postcondition $\lambda s\ldot \thelock_s(s) = \nown$
of $\mathsf{unlock}$ in Section~\ref{sec:overview} is stable, because
other-stepping cannot change the \emph{self}-component
$\thelock_s$. On the other hand, the predicate $\lambda
s\ldot \unlock(s)$ is not stable, as already commented in
Section~\ref{sec:overview}, because the value of $\unlock$ can be
changed by a thread other-stepping by
$\dotof{\Spin}{\mathsf{set\_tr}}$.


\begin{definition}[Atomic action]\label{def:action}
Let $V$ be a resource and $A$ a type. An atomic action (or action, for
short) $a$ of type $A$, over resource $V$ is relation between a value
$v : A$, and $V$-states $s$ and $s'$, with the properties below. We
write $a\ v\ s\ s'$ to relate the values and say that $a$ executed in
input state $s$, and produced output state $s'$ and return value
$v$. The properties of $a$ are:
\begin{enumerate}
\item \emph{(internality)} for every $v$, the relation 
$a\ v$ on states is an internal transition of $V$
\item \emph{(functionality}) $v$ is uniquely determined by $s$,
  i.e., if $a\ v_1\ s\ s'_1$ and $a\ v_2\ s\ s'_2$, then $v_1 = v_2$
\end{enumerate}
In an action $a$, $s'$ is also uniquely determined by $s$, because for
each $v$, the transition $a\ v$ is functional
(Def.~\ref{def:trans}.(\ref{def:func})).
\end{definition}

We can now formally define the key concept that enables program reuse
by lifting: morphisms.

\begin{definition}[Morphism]\label{def:morph}
Let $V$ and $W$ be resources. A morphism $f : V \rightarrow W$
consists of two components:
\begin{itemize}
\item A relation on states $s_v \in \Sof{V}$ and $s_w \in \Sof{W}$, written $(s_v, s_w) \in f$
\item A function on internal transitions $f : \intof{V} \rightarrow \intof{W}$. 
\end{itemize}
The components satisfy the following properties:
\begin{enumerate}
\item\label{morph:simWV} \emph{($W$ simulates $V$ by internal steps)}
  if $t \in \intof{V}$ and $t\ s_v\ s'_v$ and $(s_v, s_w) \in f$, then
  there exists $s'_w$ such that $f(t)\ s_w\ s'_w$ and $(s'_v, s'_w)
  \in f$.
\item\label{morph:contra} \emph{(functionality)} if $(s_{v1}, s_w)
  \in f$ and $(s_{v2}, s_w) \in f$, then $s_{v1} = s_{v2}$.
\item\label{morph:simVW} \emph{($V$ simulates $W$ by other steps)} if
  $s_w \osteps[W]{} s'_w$ and $(s_v, s_w) \in f$, then there exists
  $s'_v$ such that $s_v \osteps[V]{} s'_v$ and $(s'_v, s'_w) \in f$.
\item\label{morph:frame} \emph{(frame preservation)} there exists
  function $\phi : U_W \rightarrow U_V$ (notice the contravariance),
  such that: if $(s_v, s_w \zig p) \in f$, then $s_v = s'_v \zig \phi\
  p$, and $(s'_v \zag \phi\ p, s_w \zag p) \in f$.
\item\label{morph:other} \emph{(other-fixity)} if $(s_v, s_w) \in f$
  and $(s'_v, s'_w) \in f$ and $a_o(s_w) = a_o(s'_w)$ then $a_o(s_v) =
  a_o(s'_v)$.
\end{enumerate}
\end{definition}

Property~(\ref{morph:simWV}) is a relatively standard statement of
simulation: whenever $V$ can make a step by some (internal) transition
$t$ to move from $s_v$ to $s'_v$, then $W$ can follow. That is, $W$
can transition from a state $s_w$ into $s'_w$. Moreover, it is
required that $(s_v, s_w) \in f$ and $(s'_v, s'_w) \in f$. The
matching step of $W$ is constructively computed by $f$'s transition
component, in order to support program lifting in rule $\lift$ of
Section~\ref{sec:intro}, i.e., the on-the-fly modification of $e$ in
$V$ to $\mathsf{morph}~f~e$ in $W$.


Functionality property~(\ref{morph:contra}) requires that $f$, when
viewed as a relation on states, is a partial function from $W$ to $V$
(note the contravariance). This property is essential for the
soundness of the $\lift$ rule. The lifting, formally defined in
Appendix~\ref{sec:model}, logically functions as follows: it takes a
state $s_w \in S_w$, transforms it into $s_v \in S_v$ by applying the
state component of $f$, then simulates $e$'s transitions, by $f$,
starting from $s_v$, to compute the corresponding modification to
$s_w$. Functionality ensures that $s_v$ is uniquely determined from
$s_w$, as otherwise we would not know precisely in which $V$ state to
start the simulated execution of $e$. 


Functionality may look restrictive at the moment, as the customary
definitions of simulation in the literature require the state
component to be a relation, not necessarily a function. However, the
property is required by the specifics of our setting. In the
literature, simulations are usually considered between STSs that
themselves typically represent some kind of programs. For us, the STSs
are part of the program's type, and we consider how the simulation
affects the program, not just the type. The additional level of
consideration imposes the additional property. Nevertheless, we show
in Section~\ref{sec:param} that the restriction can be lifted by a
relatively simple generalization to \emph{indexed morphism families}.

Property~(\ref{morph:simVW}) states a simulation in the opposite
direction, i.e., $V$ simulates $W$, but using the reflexive-transitive
closure of other-stepping. 
Intuitively, the property ensures that we may view the interference in
$W$ as interference in $V$. Thus, a morphism $f$ actually consists of
two simulations, which work in opposite directions, but whose
definitions are very different.
In particular, the simulation in property~(\ref{morph:simVW}) only
depends on $f$'s state component, and, unlike the simulation in
property~(\ref{morph:simWV}), it is not given constructively by $f$'s
transition component. For example, in Section~\ref{sec:morph}, one may
see that $\Spin$ simulates $\CSL$ in the sense of
property~(\ref{morph:simVW}), because each transition in $\CSL$
is a coupling of a transition in $\Spin$. 
The reason for the difference between the two simulations is 
%
that the simulation in property~(\ref{morph:simVW}) is not used to
modify programs on the fly, but merely to ensure the soundness of the
$\lift$ rule. The premiss of $\lift$ specifies $e$ only under the
assumption that the interfering threads respect
$V$. $\mathsf{Morph}~f~e$ logically executes $e$, modifying its
transitions by $f$, as described above. Thus, unless we can view
interference to $\mathsf{morph}~f~e$ in $W$ as interference to $e$ in
$V$, we cannot use the specification of $e$ to infer anything about
$\mathsf{morph}~f~e$.

Properties~(\ref{morph:frame}) and (\ref{morph:other}) state
preservation of the subjective structure between the states of $V$ and
$W$. Property~(\ref{morph:frame}) says that whenever we frame by $p$
in $W$, there is a uniquely determined frame $\phi~p$ in $V$ that
corresponds to it. For example, in the case of the morphism $f
: \Spin \rightarrow \CSL$ in Section~\ref{sec:morph}, $\dotof{U}{\CSL}
= \dotof{U}{\Spin} \times \dotof{U}{\Xfer}$, and $\phi$ is defined as
the first projection, following the definition of $f$'s state
component.
%
Property~(\ref{morph:other}) requires that the \emph{other} fields are
preserved by $f$. When $f$ maps $s_w$ to $s_v$, then $a_o(s_w)$ only
depends on $a_o(s_v)$, but not on $a_s(s_w)$ and $a_j(s_w)$.


We close the section with the definition of $f$-stepping (i.e.,
stepping under a morphism $f:V \rightarrow W$), and its associated
property of $f$-stability. These are similar to \emph{other}-stepping
and stability (Definitions~\ref{def:otherstep} and~\ref{def:stable}),
but where the latter consider interference of other threads,
$f$-stepping considers steps that are $f$-images of internal
transitions of $V$.  Intuitively, $f$-stable predicates are preserved
by programs morphed by $f$. For example, in the $\lift$ rule in
Section~\ref{sec:intro}, the morphism $f : V \rightarrow W$ lifts the
program $e$, and preserves the $f$-stable predicate $I$. In
Section~\ref{sec:morph}, the predicate
$I \eqdef \lambda s\ldot \theheap_s(s) = h$ used to lift
$\mathsf{lock}$ to $\mathsf{lock'}$ is stable under morphisms
$f : \Spin \rightarrow \CSL$, because the images under $f$ of internal
transitions of $\Spin$ do not modify the \emph{self} heap in $\CSL$.

\begin{definition}[$f$-stepping]
Let $f : V \rightarrow W$ be a morphism, and $s_w$, $s'_w$ be
$W$-states. We say that $s_w$ steps by $f$ to $s'_w$, written
$s_w \mstep[f]{} s'_w$, if one of the following is true: 
\begin{enumerate}
\item there exists $t \in \intof{V}$ and $s_v$, $s'_v$, such that $(s_v, s_w) \in f$,
  $(s'_v, s'_w) \in f$, $t\ s_v\ s'_v$ and $f(t)\ s_w\ s'_w$ 
\item $s_w \ostep[W]{} s'_w$
\end{enumerate}
In other words, either $s_w$ steps into $s'_w$ by interference on $W$,
or the step is an $f$-image of an internal transition in $V$.  We
write $\msteps[f]{}$ for reflexive-transitive closure of
$\mstep[f]{}$.
\end{definition}

\begin{definition}[$f$-stability]
Let $f : V \rightarrow W$ be a morphism. Predicate $P$ over $W$-states
is \emph{$f$-stable in state $s$} if whenever $s
\msteps[f]{} s'$, then $P\ s'$. $P$ is $f$-stable if it is $f$-stable in state $s$ for every $s$ for which $P\ s$. 
%
Given $P$, we define its
\emph{$f$-stabilization} $\fstab{P}{f}$ as
$
\fstab{P}{f}\ s \eqdef \forall s'\ldot s \msteps[f]{} s' \rightarrow P\ s'.
$
It is easy to see that $\fstab{P}{f}$ is $f$-stable, and that $P$ is
$f$-stable iff $\forall s\ldot P\ s \rightarrow \fstab{P}{f}\ s$.
\end{definition}



\subsection{Basic constructions}\label{sec:constructions}

\begin{definition}[Identity and composition]
The \emph{identity morphism} $1_V$ on a resource $V$ consists of the
following state and transition components:
\begin{itemize}
\item $(s, s') \in 1_V$ iff $s = s'$
\item for every $t \in \intof{V}$, $1_V(t) = t$
\end{itemize}
Let $f : V \rightarrow W$ and $g : W \rightarrow X$ be morphism. The
composition morphism $g \circ f : V \rightarrow X$ consists of the
following state and transition components:
\begin{itemize}
\item $(s, s') \in g \circ f$ iff there exists $s''$ such that $(s,
  s'') \in f$ and $(s'', s') \in g$.
\item for every $t \in \intof{V}$, $(g \circ f)(t) = g(f(t))$
\end{itemize}
It is easy to show that 
$\circ$ is associative, with $1_V$ (resp.~$1_W$) as the right
(resp.~left) identity.
\end{definition}

\begin{definition}[Resource restriction]\label{def:resstrength}
Let $V$ be a resource, and $I$ a global $V$-inductive
predicate. Restriction of $V$ by $I$, denoted $V/I$, is a resource
defined over the same PCM and type as $V$, and with state space,
flattening, and transitions defined as follows, to make $I$ hold
constantly.
\begin{enumerate}
\item $\Sof{(V/I)}(s) \eqdef \Sof{V}(s) \wedge I(s)$
\item $\Flatof{(V/I)}{s} \eqdef \Flatof{V}{s}$
\item $\intof{(V/I)} = \intof{V}$
\item $t \in \extof{(V/I)}$ if there exists $t' \in \extof{V}$ such that
  $t\ s\ s'$ iff $t'\ s\ s' \wedge I\ s'$.
\end{enumerate}
There is a generic morphism from $V$ to $V/I$, which is identity on
states and transitions.
\end{definition}

In (1), we conjoin $I$ as an additional property to the state space of
$V$. We require that $I$ is global, so that $\Sof{(V/I)}$ is global
too, as required by Definition~\ref{def:coh}. Conditions (2-3)
propagate the flattening function and internal transitions from
$V$. Because $I$ is inductive, the internal transitions preserve
$\Sof{(V/I)}$, as required by Definition~\ref{def:resource}.  Finally,
Condition (4) strengthens the external transitions of $V$; it requires
that in $V/I$, an external transition can only be taken if it
preserves $I$.
The frequent use of restriction is to rule out undesired states from
resource composition. We will illustrate 
this in
Section~\ref{sec:rwlock}, where the functionality of readers and
writers is composed into a resource for readers/writers lock. Because
there is a dependence between the individual resources for readers and
for writers, restriction will be used to remove some state pairs from the composition.


\subsection{Inference rules}\label{sec:rules}
The inference rules of \FCSL differentiate between two different
program types: $\mathsf{ST}\ V\ A$ and
$[\Gamma]\ldot \{P\}\ A\ \{Q\}@V$. The first type encompasses programs
that respect the transitions of the resource $V$, and return a value
of type $A$ if they terminate. The second type is a subset of
$\mathsf{ST}\ V\ A$, selecting only those programs that satisfy the
precondition $P$ and postcondition $Q$. Here, $\Gamma$ is a context of
specification-only variables that serve to relate pre- and
post-states, as illustrated in Section~\ref{sec:overview}. $P$ and $Q$
are predicates drawn from the Calculus of Inductive Constructions
(CiC) which is the logic of Coq, and $A$ is a type in CiC.

The key concept in the inference rules is a predicate transformer
$\vrf~e~Q$, which takes a program $e : \mathsf{ST}\ V\ A$, and
postcondition $Q$, and returns the set of $V$-states from which $e$ is
safe to run, and produces an ending state and result result satisfying
$Q$ (thus, technically, $Q :
A \rightarrow \mbox{$V$-state} \rightarrow \mathsf{prop}$). $\mathsf{Vrf}$
is used to encode via Hoare triple types that $e$ has a precondition
$P$ and postcondition $Q$.\footnote{We abstract current state as
customary in separation logic. Otherwise, the definition reads
$\forall \Gamma\ldot \Sof{V}~s\rightarrow P~s\rightarrow \wpeqs e Q
s$.}
\[
[\Gamma]\ldot \{P\}\ A\ \{Q\}@V = \{ e : \mathsf{ST}\ V\ A \mid \forall \Gamma\ldot \Sof{V}\barrow P \barrow \wpeq e Q\}
\]

In Appendix~\ref{sec:model}, we define the denotational semantics in
CiC for $\mathsf{ST}\ V\ A$, and define the $\vrf$ predicate
transformer. Thus, we can use Coq as our environment logic, and
combine the Hoare triple types with other type constructors, to form
higher-order computations. Here we just mention that we can now
immediately give the following type to the fixed-point combinator,
where $T$ is the dependent type $T =
\Pi_{x:A}\ldot[\Gamma]\ldot\{P\}\ B\ \{Q\}@ V$ of functions of
argument $x : A$ producing concurrent computation with precondition
$P$ and postcondition $Q$:
\[
\mathsf{fix} : (T \rightarrow T) \rightarrow T.
\]
$T$ serves as a loop invariant; in $\mathsf{fix}~(\lambda f\ldot e)$
we assume that $T$ holds of $f$, but then have to prove that it holds
of $e$ as well, i.e., it is preserved upon the end of the iteration.

In the actual reasoning about programs, we keep the predicate
transformer $\vrf$ abstract, and only rely on the following minimal
set of lemmas, all proved in Coq, and presented here in separation
logic notation to implicitly abstract over the current state. These,
together with the typing for $\mathsf{fix}$ above, are the only
Hoare-related rules of \FCSL, though, of course,
\FCSL also inherits all the inference rules of CiC.
\[
\begin{array}{rcl}
\mathsf{\vrf\_vs} & : & \wpeq e Q \barrow \Sof{V}\\
\mathsf{\vrf\_post} & : & (\forall r~s\ldot \Sof{V}~s \rightarrow Q_1~r~s \rightarrow Q_2~r~s) \barrow
                \wpeq e {Q_1} \barrow \wpeq e {Q_2}\\
\mathsf{\vrf\_ret} & : & 
    \Sof{V} \barrow \stab{(Q\ r)} \barrow \wpeq {(\mathsf{ret}\ r)} Q\\
\mathsf{\vrf\_bnd} & : & 
   \wpeq {e_1} {(\lambda x\ldot \wpeq {(e_2\ x)} Q)} \barrow
   \wpeq {(x\leftarrow e_1; (e_2\ x))} Q\\
\mathsf{\vrf\_par} & : & 
  (\wpeq {e_1} {Q_1}) \bstar (\wpeq {e_2} {Q_2}) \barrow
  \wpeq {(e_1 \parallel e_2)} {(\lambda r:A_1\times A_2\ldot (Q_1\ r.1)\ {\bstar}\ (Q_2\ r.2))}\\
\mathsf{\vrf\_frame} & : & 
 (\wpeq {e} {Q_1}) \bstar {\stab Q_2} \barrow \wpeq {e} {(\lambda r\ldot (Q_1\ r) \bstar Q_2)}\\
\mathsf{\vrf\_atm} & : & 
   \Sof{V} \barrow \stab{(\lambda s\ldot \exists r\ s'\ldot a\ r\ s\ s' \wedge \stab{(Q\ r)}\ s')} \barrow 
    \wpeqs {(\mathsf{atomic}\ a)} Q\\
\mathsf{\vrf\_morph} & : & 
    \smorph f {(\wpeq e Q)} \bwedge \fstab{I}{f} \barrow
    \wpeqs {(\mathsf{morph}\ f\ e)} {(\lambda r\ldot \smorph f {(Q\ r)} \bwedge I)}\\
\end{array}
\]
%
The $\mathsf{\vrf\_vs}$ lemma says that if a state is in $\wpeq e Q$,
then it is also in $V$'s state space. In other words, the predicate
transformer $\vrf$ is only concerned with states that are valid for
the resource $V$.

The $\mathsf{\vrf\_post}$ lemma says that we can weaken the
postcondition $Q_1$ into $Q_2$ if the first implies the second for
every return value $r$ and state $s$. The lemma is thus a variant of
the customary Hoare logic rule of consequence. When proving $Q_2$ out
of $Q_1$, it is sound to further assume $\Sof{V}$, because $\vrf$ is
only concerned with states that are valid for the resource $V$.



The $\mathsf{\vrf\_ret}$ lemma states that if $Q\ r$ holds in the
initial states, then the ending state of $\mathsf{ret\ r}$ satisfies
$Q\ r$; in other words, $\mathsf{ret}\ r$ does not change the state
and just returns $r$. To account for the possibility that the
environment threads may change the state, we stabilize $Q\ r$ in the
premiss.

The $\mathsf{vrf\_bnd}$ lemma is the customary Dijkstra-style rule for
computing a predicate transformer of a sequential composition, by
nesting two applications of the transformer.

The $\mathsf{\vrf\_par}$ lemma encodes the usual property of
separation logics that if the initial state $s$ can be split into
$s_1$ and $s_2$, such that $e_1$ executes in $s_1$ to obtain
postcondition $Q_1$, and $e_2$ executes in $s_2$ to obtain
postcondition $Q_2$, then the ending state of $e_1 \parallel e_2$ can
be split in the same way.
%
%
%
This follows from the definition of $P
\bstar Q$ which is slightly different than in separation
logic, to account for \FCSL's different notion of state.
\[
(P \bstar Q)\ s  \eqdef \exists x_1\ x_2\ldot a_s(s) = x_1 \join x_2 \wedge
                            P\ (x_1, a_j(s), a_o(s) \join x_2) \wedge 
                            Q\ (x_2, a_j(s), a_o(s) \join x_1).
\]
The definition captures the state view of the children threads $e_1$
and $e_2$ upon their forking in the parent state
$s$. The \emph{self}-components of the children states divide
the \emph{self}-component of the parent ($a_s(s) = x_1 \join x_2$). At
the same time, the \emph{other}-component of $e_1$ adds
the \emph{self}-components of $e_2$ ($a_o(s) \join x_2$) to capture
the fact that $e_2$ becomes part of the concurrent environment of
$e_1$, and vice versa. The joint component $a_j(s)$ represents shared
state, so it is propagated to both children without changing.
Finally, the end-result of $e_1 \parallel e_2$ is a pair $r = (r.1,
r.2)$ of type $A_1 \times A_2$, combining the return results of $e_1$
and $e_2$, of types $A_1$ and $A_2$, respectively. Thus, the
postcondition of $e_1 \parallel e_2$ splits $r$ and passes the
projections to $Q_1$ and $Q_2$.

The $\mathsf{\vrf\_frame}$ lemma is, intuitively, a form of
$\mathsf{\vrf\_par}$ lemma where $e_2$ is taken to be an idle
thread. Thus, it can be seen as a combination of $\mathsf{\vrf\_par}$
and $\mathsf{vrf\_ret}$ lemmas, which is why we stabilize $Q_2$ in the
premiss.

The $\mathsf{\vrf\_atm}$ lemma says $Q$ is a postcondition for an
action $a$ in the pre-state $s$, if there exist the return value $r$
and post-state $s'$ that are related by $a$ (i.e., such that $a\ r\ s\
s'$) and $Q~r~s'$. We allow for environment steps before $s$ and after
$s'$, which is why we stabilize the whole predicate binding $s$, and
we stabilize $Q\ r$ before applying it to $s'$.

Finally, $\mathsf{\vrf\_morph}$ is a predicate-transformer version of
$\lift$ rule from Section~\ref{sec:intro}.\footnote{Indeed, the latter
is a direct consequence of $\mathsf{\vrf\_morph}$ and the definition
of Hoare triple type.} Unfolding the definition of $\smorph{f}{P}
= \lambda s_w.\, \exists s_v.\, (s_v, s_w) \in f \wedge P\ s_v$, the
lemma says that if we are given the initial $W$-state $s_w$, for which
there exists $s_v$ such that $(s_v, s_w) \in f$, and if running $e$ in
$s_v$ results in the postcondition $Q$, then running $\mathsf{morph}\
f\ e$ in $s_w$ will first switch to $s_v$, execute $e$ there, and then
come back to obtain the ending state satisfying $\smorph{f}{Q}$. The
predicate $I$ is propagated from the premiss to the conclusion, but is
stabilized in the pre-state to avoid the side-condition that $I$ is
$f$-stable.



\section{Readers/Writers}\label{sec:rwlock}
\newcommand{\RWLock}{\mathsf{RWLock}}
\newcommand{\RLock}{\mathsf{RLock}}
\newcommand{\WLock}{\mathsf{WLock}}
\newcommand{\Count}{\mathsf{Count}}
\newcommand{\lk}{\lambda}
\newcommand{\counter}{\mathsf{ct}}
\newcommand{\thecount}{\kappa}
\newcommand{\incr}{\iota}
\newcommand{\prologue}{\mathsf{prologue}}
\newcommand{\epilogue}{\mathsf{epilogue}}

This section illustrates component reuse on the example of
readers-writers locks~\cite{courtois:ACM71,Bornat-al:POPL05}, a
significantly more involved construction than $\CSL$ from
Section~\ref{sec:overview}.
%
The writers lock $wr$ protects a shared heap, just as in the case of
$\CSL$. When a writer acquires $wr$, it gains exclusive ownership of
the heap. But when a reader acquires $wr$, the heap becomes shared by
all concurrent readers, while becoming inaccessible to writers. To
support this discipline, the readers have to register
(resp. deregister) themselves, by incrementing (resp. decrementing) a
shared counter $\counter$ that keeps the overall number of
readers. The counter $\counter$ is protected by another lock $rd$, as
shown by the $\prologue$ (resp. $\epilogue$) procedure below.
\[
\begin{array}{cc}
\begin{array}{l}
\mathsf{prologue}() = \hbox{}\\
\quad  \mathsf{lock}(rd);\\
\quad  x\ {\leftarrow}\ ! \counter;\\
\quad  \mathsf{if}\ x = 0\ \mathsf{then}\ \mathsf{lock}(wr);\\
\quad  \counter := x+1;\\
\quad  \mathsf{unlock}(rd)
\end{array}
&
\begin{array}{l}
\mathsf{epilogue}() = \hbox{}\\
\quad  \mathsf{lock}(rd);\\
\quad  x\ {\leftarrow}\ ! \counter;\\
\quad  \counter := x - 1;\\
\quad  \mathsf{if}\ x = 1\ \mathsf{then}\ \mathsf{unlock}(wr);\\
\quad  \mathsf{unlock}(rd)
\end{array}
\end{array}
\]
The first reader to execute $\prologue$ is responsible for acquiring
$wr$, and the last reader to execute $\epilogue$ releases it, to let
the writers in. Moreover, $\epilogue$ should only be invoked by a
reader that already went through $\prologue$. Between calls to
$\prologue$ and $\epilogue$, the reader can freely read from the
shared heap, which is guaranteed not to be changed by a writer.
A thread may invoke $\prologue$ and register as a reader multiple
times. The extra registrations are not extraneous as, upon forking,
they are divided between the thread's children.  Thus, a thread
holding more than one registration is simply pre-registering its
children as readers.

From the logical standpoint, $\prologue$ and $\epilogue$ manage the
ownership of the protected heap, just as $\CSL$ did, but here the
ownership discipline is much more involved.
Intuitively, we have two distinct resources: $\WLock$ for writers, and
$\RLock$ for readers. When the heap is in the shared state of
$\WLock$, it can be acquired by a writer and moved to the writer's
private state. We say that the heap is then in ``write'' mode. This is
the functionality we already saw in $\CSL$. But here, the heap can
also be acquired by the first reader that goes through $\prologue$, in
which case the heap moves to the shared state of $\RLock$, where it
can be accessed by any reader. We say that the heap is in ``read''
mode. Dually, $\epilogue$ returns the heap from the shared state of
$\RLock$ to the shared state of $\WLock$, when invoked by the last
reader.

We can thus divide the readers/writers construction into several
sub-components. First, we formalize the two different ownership modes
by a new 
resource $\Spin'$. $\Spin'$ will implement the ``write''
mode, similar to $\Spin$ in Section~\ref{sec:overview}, but will also
enable the ``read'' mode to be added by composition with other
resources. Second, we formalize the discipline of reader registration
and deregistration, and ensure that the protected heap is in ``read''
mode if a registered reader exists. Finally, we formalize the transfer
of the protected heap between different ownership modes, by composing
instances of the resources $\Shar$ and $\Priv$ that we already
introduced in Section~\ref{sec:overview}. Ultimately, the pieces
combine into the resource $\RWLock$ for readers/writers, as
schematically illustrated in Figure~\ref{fig:rwlock}. We will explain
the figure in detail further in this section; for now, it suffices to
note that the construction instantiates each of $\Spin'$ and $\Shar$
twice (once for writers, once for readers), thus achieving
reuse. 


\begin{figure}[t]
{\small
\[
\begin{array}{|l||c|c|c||c|c|c|}
\hline
  & \multicolumn{6}{|c|}{\RWLock}\\
\hline
  & \multicolumn{3}{|c||}{\WLock} & \multicolumn{3}{c|}{\RLock}\\
\hline 
  & \Spin'(wr) & \Shar & \Priv & \Spin'_2(rd) & \Count & \Shar_2\\
\hline
  & \thelock_s, \unlock, \lk, \thelock_o & \theheap_j, \here & \theheap_s, \theheap_o & \thelock_{s2}, \unlock_2, \lk_2, \thelock_{o2} & \thecount_s, \incr, \thecount_o & \theheap_{j2}, \here_2\\
\hline
\mathsf{wrlock\_tr} & \mathsf{lock\_tr}(\own) & - & - & - & - & - \\
\hline
\mathsf{wrunlock\_tr} & \mathsf{unlock\_tr}(\own) & - & - & - & - & - \\
\hline
\mathsf{freeze\_tr} & \mathsf{lock\_tr}(\nown) & - & - & \mathsf{id\_tr}(\lambda s\ldot \thelock_{s2}(s)=\own) & - & - \\
\hline
\mathsf{unfreeze\_tr} & \mathsf{unlock\_tr}(\nown) & - & - & \mathsf{id\_tr}(\lambda s\ldot \thelock_{s2}(s)=\own) & - & - \\
\hline
\mathsf{rdlock\_tr} & - & - & - & \mathsf{lock\_tr}(\own) & - & -\\
\hline
\mathsf{rdunlock\_tr} & - & - & - & \mathsf{unlock\_tr}(\own) & -  & -  \\
\hline
\mathsf{incr\_tr} & - & - & - & \mathsf{id\_tr}(\lambda s\ldot \thelock_{s2}(s) = \own) & \mathsf{incr\_tr} & - \\
\hline
\mathsf{decr\_tr} & - & - & - & \mathsf{id\_tr}(\lambda s\ldot \thelock_{s2}(s) = \own) & \mathsf{decr\_tr} & - \\
\hline
\mathsf{open\_tr} & \mathsf{set\_tr}(\own)(\mathsf{false}) & \mathsf{give\_tr} & \mathsf{take\_tr} & - & - & - \\
\hline
\mathsf{close\_tr} & \mathsf{set\_tr}(\own)(\mathsf{true}) & \mathsf{take\_tr} & \mathsf{give\_tr} & - & - & - \\
\hline
\mathsf{toreader\_tr} & \mathsf{set\_tr}(\nown)(\mathsf{false}) & \mathsf{give\_tr} & - & \mathsf{id\_tr}(\lambda s\ldot \thelock_{s2}(s)=\own) & \mathsf{set\_tr}(\mathsf{true}) & \mathsf{take\_tr}\\
\hline
\mathsf{towriter\_tr} & \mathsf{set\_tr}(\nown)(\mathsf{true}) & \mathsf{take\_tr} & - & \mathsf{id\_tr}(\lambda s\ldot \thelock_{s2}(s)=\own) & \mathsf{set\_tr}(\mathsf{false}) & \mathsf{give\_tr}\\
\hline
\end{array}
\]}
\caption{Coupling of the transitions of $\RWLock$. The rows are the
  transitions of $\RWLock$, and the columns are the transitions of
  individual components which are coupled to provide the $\RWLock$
  transition. The top row of each column lists the fields of the
  component's state space. Empty cells indicate the $\mathsf{id\_tr}$
  transition. All the transitions of $\RWLock$ are internal.}\label{fig:rwlock}
\end{figure}

Our description of $\RWLock$ will focus on the 
%
$\prologue$ and $\epilogue$ procedures, to which we ascribe the
following specifications.
\footnote{The specifications can be simplified by taking
  $h=\mathsf{empty}$ and $c=0$; the general case can be recovered by framing.}
\[
\begin{array}{c}
\mathsf{prologue} : \specK{[h, c]}\ldot \spec{\lambda s\ldot \theheap_s(s) = h \wedge \thecount_s(s) = c}~
                                \spec{\lambda s\ldot \theheap_s(s) = h \wedge \thecount_s(s) = c+1}\specK{@\RWLock}\\
\mathsf{epilogue} : \specK{[h, c]}\ldot \spec{\lambda s\ldot \theheap_s(s) = h \wedge \thecount_s(s) = c+1}~
                                \spec{\lambda s\ldot \theheap_s(s) = h \wedge \thecount_s(s) = c}\specK{@\RWLock}
\end{array}
\]
In the specifications, $\theheap_s$ stands for the private heap of the
invoking thread, and $\thecount_s$ is the number of readers that the
thread has registered. The registration count is increased by
$\prologue$ and decreased by $\epilogue$. A thread is a reader if its
$\thecount_s(s) > 0$. Notice that $\thecount_s(s)$ is a
\emph{self}-field, which has two important consequences. First, as
described in Section~\ref{sec:formal}, the thread's value of
$\thecount_s(s)$ is divided upon forking between the thread's
children, which thereby inherit any extra registrations that the
parent may have had.
Second, if a thread is a reader, i.e.,~$\thecount_s(s) > 0$, then it
remains so under interference, as $\thecount_s(s)$ cannot be changed
by other threads. A thread can stop being a reader only if it
deregisters itself by invoking $\epilogue$.
%
%
%


\subsection{The resource $\Spin'(r)$ for locking $r$ without exclusive ownership of $r$}\label{sec:rwspin}
The $\Spin'(r)$ resource implements spin locks, but with two different
modes of ownership: exclusive ownership by the locking thread, and
non-exclusive ownership. In the instance $\Spin'(wr)$ used by
$\WLock$, exclusive ownership is used when the writer takes the writer
lock (the ``write'' mode of the heap), and non-exclusive ownership is
used when the reader takes the writer lock (the ``read'' mode): in
the latter case, the heap collectively must be owned by all the
readers. In the instance $\Spin'(rd)$ used by $\RLock$, exclusive
ownership is used when the reader takes the reader lock, while
non-exclusive ownership is not needed. 

Omitting $r$ from now on, the states of $\Spin'$ have the form $s =
(\thelock_s, (\lk, \unlock), \thelock_o)$. The boolean $\lk$ is
$\mathsf{true}$ if the underlying lock is taken, and is
$\mathsf{false}$ otherwise. As in $\Spin$, $\unlock$ is a boolean that
has to be set before unlocking; $\thelock_s, \thelock_o \in O$
indicate the exclusive ownership of the lock; and $\thelock(s) =
\thelock_s(s) \join \thelock_o(s)$.
%
%
\[
\begin{array}{rcl}
S(s) & \eqdef & \valid\ (\thelock(s)) \wedge r \neq \mathsf{null} \wedge 
    (\neg{\lk(s)} \rightarrow \thelock(s) = \nown \wedge \unlock(s))\\
\hflat{s} & \eqdef & r \hmapsto \lk(s)
\end{array}
\]
The state space imposes the condition that if the (readers or writers)
lock is free ($\neg \lk(s)$), then no thread owns the lock exclusively
($\thelock(s)=\nown$). However, \emph{it does not} impose the
implication in the other direction: it may be that the lock is taken
\emph{and} $\thelock(s) = \nown$, which models the non-exclusive
ownership.
%
%
Additionally, if the lock is free, then $\unlock(s)$; that is, the
shared heap will satisfy the invariant in the eventual composition
with a resource for heap transfer, just like in $\Spin$.

The transitions are similar to $\Spin$, except they now use $\lk(s)$
to express the lock's status, and they have to deal with two different
ownership modes. We capture the latter by adding an extra parameter $x
\in O$ to all non-idle transitions. Passing $x = \own$
(resp.~$x=\nown$) gives us the transition dealing with exclusive
(resp.~non-exclusive) ownership.
\[
\begin{array}{rcl}
\mathsf{lock\_tr}\ x\ \ s\ s' & \eqdef & \neg \lk(s) \wedge \lk(s') \wedge \thelock_s(s') = x \wedge \unlock(s')\\
\mathsf{unlock\_tr}\ x\ s\ s' & \eqdef & \lk(s) \wedge \thelock_s(s) = x \wedge \thelock_o(s) = \nown \wedge \unlock(s) \wedge
                                          \neg \lk(s')\\
\mathsf{set\_tr}\ x\ b\ s\ s' & \eqdef & \lk(s) \wedge \thelock_s(s) = x \wedge \thelock_o(s) = \nown \wedge \lk(s') \wedge \thelock_s(s') = x \wedge \unlock(s') = b\\
\end{array}
\]
For example, $\mathsf{lock\_tr}$ switches $\lk$ from $\mathsf{false}$
to $\mathsf{true}$, as one would expect. As in $\Spin$, it also sets
$\unlock(s')$. But, if invoked with $x = \own$, it also sets
$\thelock_s(s')$ to $\own$ to signal the exclusive ownership of the
lock. 
Similarly, $\mathsf{unlock\_tr}$ switches $\lk$ from $\mathsf{true}$
to $\mathsf{false}$, and also requires $\unlock(s)$ to be set, as in
$\Spin$. If invoked with $x = \own$ it requires that the invoking
thread actually has exclusive ownership of the lock. Otherwise, if
invoked with $x = \nown$, no thread is allowed to have exclusive
ownership ($\thelock_s(s)=\thelock_o(s)=\nown$). The transitions
obtained for different values of $x$ will be coupled differently in
the eventual composition. Importantly, the $x = \own$ versions of the
transitions are \internal, whereas those obtained with $x = \nown$ are
\external, as the notion of ownership that the latter represents will
be formalized only when we compose with the resource for readers.
The $\mathsf{set\_tr}\ x\ b$ transition sets $\unlock(s')$ to $b$. It
requires the lock to be held ($\lk(s)$), but not exclusively by other
threads ($\thelock_o(s) = \nown$). Thus, in the composition, $\unlock$
could be changed by any reader, if the readers have acquired the
writer lock, but only by the writer that owns the lock.
It may be interesting to observe here that passing $x = \own$ to the
transitions essentially recovers the functionality of $\Spin$, whereas
passing $x = \nown$ produces new transitions. If we strengthen the
state space of $\Spin'$ to include $\lk(s) \rightarrow
\thelock(s)=\own$, then none of the new transitions can ever be
invoked, because the conditions on their initial state will never be
satisfiable. Thus, $\Spin'$ reduces to $\Spin$, when $x = \own$.
%

\subsection{The counting resource $\Count$}\label{sec:count}
The resource $\Count$ tracks reader registration. Physically, the
registration count is kept in the pointer $\counter$, but it is the
division of the count into \emph{self} and \emph{other} fields that is
important for the specification of $\prologue$ and $\epilogue$.  The
states of $\Count$ thus have the form $s = (\thecount_s, \incr,
\thecount_o)$, where $\thecount_s$ and $\thecount_o$ keep the number
of registrations made by the invoking thread and its environment,
respectively. In every resource, the \emph{self} and \emph{other}
components must be drawn from the PCM; here it is the PCM of natural
numbers under $+$, with $0$ as the unit element.
%
%
The field $\incr$ is a boolean, motivated similarly to $\unlock$ in
Section~\ref{sec:overview}--it indicates in the eventual composition
of $\Count$ into $\RLock$ that the heap on which the readers are to
operate is in ``read'' mode.
%
%
The above description motivates the following state-space design for
$\Count$.
\[
\begin{array}{rcl}
S(s) & \eqdef & \counter \neq \mathsf{null} \wedge \thecount(s) > 0 \rightarrow \incr(s)\\
\hflat{s} & \eqdef & \counter \hmapsto \thecount(s)
\end{array}
\]
The conjunct $\thecount(s) > 0 \rightarrow \incr(s)$ ensures that 
if there are registered readers, then, in the composition, the heap is
in ``read'' mode. The conjunct $\counter \neq \mathsf{null}$ requires
that $\counter$ is a valid pointer. 

The non-idle transitions of $\Count$ are as follows. 
%
\[
\begin{array}{rcl}
\mathsf{incr\_tr}\ s\ s' & \eqdef & \incr(s) \wedge \thecount_s(s') = \thecount_s(s)+1\\
\mathsf{decr\_tr}\ s\ s' & \eqdef & \thecount_s(s')+1 = \thecount_s(s) \wedge \incr(s')\\
\mathsf{set\_tr}\ b\ s\ s' & \eqdef & \thecount(s) = \thecount(s') = 0 \wedge \incr(s') = b
\end{array}
\]
In English, $\mathsf{incr\_tr}$ increments $\thecount_s(s)$, but
requires that the $\incr(s)$ bit is set, that is, the heap is in
``read'' mode. Similarly, $\mathsf{decr\_tr}$ decrements
$\thecount_s(s)$, but the latter has to be non-zero---a reader can
cancel only the registration that it had made itself. By the
definition of $S$, if $\thecount_s(s)>0$ in the pre-state, then
$\incr(s)$ is set, and $\mathsf{decr\_tr}$ keeps $\incr$ set in the
post-state. If $\thecount_s(s) = 0$, then $\mathsf{decr\_tr}$ cannot
execute. $\mathsf{Set\_tr}\ b$ sets $\incr(s')$ to $b$, but it
requires (and maintains) that $\thecount(s)=0$; that is, the ownership
mode of the heap can be changed only when there are no readers in the
system.


\subsection{Composing into $\RWLock$}\label{sec:rwcomp}
We now combine the components into a resource $\RWLock$, as shown in
Figure~\ref{fig:rwlock}.
%
%
The fields of the combination contain the fields of $\WLock$, tracking
information about writers, and of $\RLock$, tracking information about
readers.
The $\WLock$ state is itself a product of the state-spaces of
$\Spin'(wr)$, $\Shar$, and $\Priv$. Here, $\Shar$ provides the
functionality of a shared heap with an invariant $I$. When the
protected heap is in this sub-resource, it is in $\WLock$, but is not
owned by any thread. $\Priv$ provides the functionality of private
heaps, with the operations for lookup, update, allocation and
deallocation, whose discussion we elide here. When the heap is in
$\Priv$, it is owned exclusively by a writer that locked it, i.e., the
heap is in the ``write'' mode.
The $\RLock$ state is a product of the state-spaces of $\Spin'(rd)$,
$\Count$ and $\Shar$. Here $\Spin'(rd)$ provides the functionality of
the spin lock $rd$. $\Shar$ provides the functionality of a shared
heap with an invariant $I$. When the protected heap is in this
sub-resource, it is in $\RLock$, and owned collectively by all
readers, that is, it is in ``read'' mode. To differentiate these instances
of $\Spin'$ and $\Shar$ from the ones used in $\WLock$, we index them
and their fields by $2$.

The state space of $\RWLock$, however, cannot be a simple product of
the underlying components, and we need to impose the additional
invariant $\mathsf{RWinv}$ defined below. Thus, we first build an
intermediate resource $\RWLock'$ which combines the states and
transitions as shown in Figure~\ref{fig:rwlock}, then construct the
restriction $\RWLock = \RWLock'/\mathsf{RWinv}$ (see
Definition~\ref{def:resstrength}), and inject $\RWLock'$ into
$\RWLock$ by the generic morphism for resource restrictions. 
\[
\begin{array}{rcl}
\mathsf{RWinv}(s) & \eqdef & \unlock(s) = \here(s) \wedge \incr(s) = \here_2(s) \wedge \unlock_2(s) \wedge 
(\lk_2(s) \rightarrow \thelock_{s2}(s) = \own) \wedge \hbox{}\\
     &        & (\here_2(s) \leftrightarrow \lk(s) \wedge \thelock(s) = \nown \wedge \neg \here(s)) 
\end{array}
\]
%
The first and second conjuncts of $\mathsf{RWinv}$ capture that
$\unlock$ in $\Spin'(wr)$ and $\incr$ in $\Count$ are proxies for the
presence of the protected heap in $\Shar$ and $\Shar_2$,
respectively. This is similar to how we equated $\unlock$ and $\here$
in the state space of $\CSL$ in Section~\ref{sec:overview}.
The third conjunct fixes the value of $\unlock_2(s)$, indicating that
we are not going to be coupling $\mathsf{unlock\_tr}$ of $\Spin'(rd)$
in non-trivial ways.
%
The fourth conjunct excludes the possibility for the collective
ownership of $rd$, as the reader lock will always be acquired
exclusively by readers.
%
Finally, the last conjunct describes the possible states in which the
protected heap may be. It says that the protected heap is in $\RLock$
($\here_2(s)$) iff the writer lock is taken ($\lk(s)$) by readers
($\thelock(s) = \nown$), and the heap is not in $\WLock$ ($\neg
\here(s)$).

The $\mathsf{wrlock\_tr}$ and $\mathsf{wrunlock\_tr}$ in
Figure~\ref{fig:rwlock} are transitions for exclusive locking and
unlocking by the writer. Thus, they lift the locking and unlocking
transition from $\Spin'(wr)$, and do so by coupling with identity
transitions across the board. We use the $\own$ version of the
transition, i.e., locking and unlocking for exclusive ownership. The
transitions $\mathsf{freeze\_tr}$ and $\mathsf{unfreeze\_tr}$
correspond to the reader locking and unlocking the writer lock,
respectively, and thus couple with the $\nown$ version of $\Spin'(wr)$
locking and unlocking transitions. They also require in $\Spin'(rd)$
that the reader lock is owned. Hence, a reader can try to lock and
unlock the writers lock, but only if she first obtains the readers
lock. 
%
We emphasize how the relationship between the various fields ensures
that $\mathsf{freeze\_tr}$ and $\mathsf{unfreeze\_tr}$ can only be
invoked when $\thecount(s) = 0$, i.e., the invoking reader is the sole
reader in the system, and has not yet incremented $\thecount(s)$
(first reader), or has just decremented $\thecount(s)$ (last
reader). Indeed, if $\thecount(s) > 0$ then $\incr(s)$ by
$\Sof{\Count}$. But then, $\here_2(s)$ by $\mathsf{RWinv}$, and then
also $\lk(s)$, $\neg\here(s)$ and $\neg\unlock(s)$.  But the
subcomponent $\mathsf{lock\_tr}(\nown)$ of $\mathsf{freeze\_tr}$
requires $\neg\lk(s)$, and the subcomponent
$\mathsf{unlock\_tr}(\nown)$ of $\mathsf{unfreeze\_tr}$ requires
$\unlock(s)$.


%
The transitions $\mathsf{rdlock\_tr}$ and $\mathsf{rdunlock\_tr}$
implement the locking and unlocking of the readers lock, and thus
invoke the respective $\own$ version of the $\Spin'(rd)$
transitions.
%
%
The $\mathsf{incr\_tr}$ and $\mathsf{decr\_tr}$ are straightforward
lifting from $\Count$, but can only be invoked in the combination by a
thread holding the reader lock.

Finally, the last four transitions implement the ownership transfer of
the heap within a $\WLock$ resource, and between $\WLock$ and
$\RLock$. Transitions $\mathsf{open\_tr}$ and $\mathsf{close\_tr}$ move the heap
between $\WLock$ shared state (when the heap is not owned by anybody)
and the writers resource private heap (``write'' mode). On the other
hand, $\mathsf{toreader\_tr}$ moves the heap from $\WLock$ to
$\RLock$, setting the heap to ``read-only'' mode. Notice how the
transition synchronizes the boolean fields $\unlock$ in $\Spin'(wr)$
and $\incr$ in $\Count$, to capture that the first is set to
$\mathsf{false}$ simultaneously with the second being set to
$\mathsf{true}$. Transition $\mathsf{towriter\_tr}$ works in the opposite
direction.

\subsection{Annotating and verifying $\mathsf{prologue}$}\label{sec:prologue}
We next present the proof outline for $\mathsf{prologue}$ in
Figure~\ref{fig:pfprologue} (the similar proof for $\mathsf{epilogue}$
is in the Coq files). In the code, we replace the physical operations
such as, e.g., reading from $\counter$ and writing into it, with
actions. Actions thus decorate the physical operations with auxiliary
code, built out of the transition of $\RWLock$, and the program erases
to the one given in Section~\ref{sec:rwlock}.

\begin{figure}
{\small
\[
\begin{array}{l}
\mathsf{prologue}() = \hbox{}\\
\begin{array}[t]{r@{\quad}l}
1. & \spec{\theheap_s(s) = h \wedge \thecount_s(s) = c}\\
2. & \act{rdlock};\\
3. & \spec{\theheap_s(s) = h \wedge \thecount_s(s) = c \wedge \thelock_{s2}(s) = \own}\\
4. & x\ {\leftarrow}\ \mathsf{atomic}~(\act{readcnt\_act});\\
5. & \spec{\theheap_s(s) = h \wedge \thecount_s(s) = c \wedge \thelock_{s2}(s) = \own \wedge x = c + \thecount_o(s)}\\
6. & \mathsf{if}\ x = 0\ \mathsf{then}\ \act{freeze}; \mathsf{atomic}~(\mathsf{toreader\_act}); \\
7. & \spec{\theheap_s(s) = h \wedge \thecount_s(s) = c \wedge \thelock_{s2}(s) = \own \wedge x = c + \thecount_o(s) \wedge \incr_s(s)}\\
8. & \mathsf{atomic}~(\act{incr\_act}\ x); \\
9. & \spec{\theheap_s(s) = h \wedge \thecount_s(s) = c+1 \wedge \thelock_{s2}(s) = \own}\\
10. & \mathsf{atomic}~(\mathsf{rdunlock\_act})\\
11. & \spec{\theheap_s(s) = h \wedge \thecount_s(s) = c+1}\vspace{-4mm}
\end{array}
\end{array}
\]}
\caption{Proof outline for $\mathsf{prologue}$.\vspace{-4mm}}\label{fig:pfprologue}
\end{figure}

In line~2, $\mathsf{rdlock}$ is a procedure that loops over the
spin-lock $\mathsf{rd}$, trying to acquire it by means of
$\mathsf{rdlock\_tr}$ transition in $\RWLock$. The latter is a
coupling of $\Spin'_2(rd).\mathsf{lock\_tr}(\own)$ with
$\mathsf{id\_tr}$ on all sub-components
(Figure~\ref{fig:rwlock}). Thus, it sets $\thelock_{s2}$ to $\own$,
preserving the other components. In particular, the values of
$\theheap_s$ and $\thecount_s$ are propagated from line~1 to
line~3. For brevity, we omit the definition of $\mathsf{rdlock}$; it
is implemented by lifting, and thus reusing, the $\mathsf{lock}$
procedure for $\Spin'$, exactly in the same way that we produced
$\mathsf{lock'}$ out of $\mathsf{lock}$ in Section~\ref{sec:overview}.

The action $\mathsf{readcnt\_act}$ is defined as follows.
\[
\mathsf{readcnt\_act}\ x\ s\ s' \eqdef \mathsf{id\_tr}\ (\lambda s\ldot \thecount(s) = x)\ s\ s'
\]
As it invokes $\mathsf{id\_tr}$, the action does not change the state,
but the predicate $\lambda s\ldot\thecount(s) = x$ ties the return
result $x$ to $\thecount(s)$, which equals the contents of
$\counter$. Thus, $\mathsf{read\_act}$ erases to a lookup of
$\counter$.

Line~6 ensures that the protected heap is acquired by the readers. If
$x > 0$, then by the state space of $\Count$, we know that
$\thecount(s) > 0$ and thus, $\incr(s)$. On the other hand, if $x =
0$, we invoke $\mathsf{freeze};
\mathsf{toreader\_act}$. $\mathsf{Freeze}$ is a locking procedure,
just like $\mathsf{rdlock}$. However, it loops over $wr$, trying to
execute the $\mathsf{freeze\_tr}$ transition, which is composed out of
$\mathsf{Spin}.\mathsf{lock\_tr}(\nown)$ with a number of idle
transitions. In the outcome, the loop terminates with $wr$ lock taken,
and $\here(s)$ field set, indicating that the protected heap is in the
writer resource. Thus, we subsequently execute
$\mathsf{toreader\_act}$ to move the heap to the reader resource, and
thus set $\here_2(s)$. As the invariant $\mathsf{RWinv}$ equates
$\here_2(s) = \incr(s)$, we know that $\incr(s)$ holds in line~7.
%
%
%
Thus, we can invoke $\act{incr\_act}\ x$, defined as:
\[
\act{incr\_act}\ x\ s\ s' \eqdef x = \thecount(s) \wedge \act{incr\_tr}\ s\ s'
\]
The action transitions by $\mathsf{incr\_tr}$ to increment
$\thecount_s(s)$. It requires $\thecount(s)$, which is the contents of
$\counter$, to equal $x$; hence, it erases to the physical operation
of writing of $x+1$ into $\counter$.
Finally, in line~10, $\mathsf{rdunlock\_act}$ invokes
$\RWLock.\mathsf{rdunlock\_tr}$ to release the $rd$ lock, giving us
the final specification. 

\newcommand{\Stack}{\mathsf{Stack}}
\newcommand{\heap}{\mathsf{heap}}
\newcommand{\hist}{\mathsf{hist}}
\newcommand{\cons}[2]{#1\,{::}\,#2}
\newcommand{\thesent}{\pi}
\newcommand{\thestack}{\alpha}
\newcommand{\islist}[2]{\mathsf{layout}\,#1\,#2}
\newcommand{\liftx}{\textsc{LiftX}\xspace}

\section{Indexed morphism families and quiescence}\label{sec:param}

As defined in Section~\ref{sec:formal}, the state component of a
morphism $f : V \rightarrow W$ is a (partial) function from $\Sof{W}$
to $\Sof{V}$. Functionality is required for $f$ to be able to lift
programs from $V$ to $W$. Indeed, given a program $e$ over $V$, and a
$W$-state $s_w$, lifting requires first mapping $s_w$ into a $V$-state
$s_v$, in order to run $e$ on $s_v$. It is only sensible for $s_v$ to
be uniquely determined by $s_w$, and we were not able to prove the
\lift rule sound without functionality.

There are examples, however, as we will show, where we would like $f$
to be a relation on states, but not a function. To 
reconcile the two contradictory requirements, we generalize morphisms to indexed
morphism families (or just families, for short), as follows.
A family $f : V \rightarrowX{X} W$ introduces a type $X$ of indices
for $f$. The state component of $f$ is a partial function $f : X
\rightarrow \Sof{W} \rightharpoonup \Sof{V}$, and the transition
component of $f$ is a function $f : X \rightarrow \transof{V}
\rightarrow \transof{W}$, satisfying a number of properties (listed in
Appendix~\ref{sec:famdef}), which reduce to properties of morphisms
when $X$ is the $\mathsf{unit}$ type.
By choosing $X$ suitably, we can represent any relation $R \subseteq
\Sof{W} \times \Sof{V}$ as a partial function $f_R : X \rightarrow
\Sof{W} \rightharpoonup \Sof{V}$. Indeed, we can take $X = \Sof{V}$,
and set $f_R~s_v~s_w = s_v$ if $(s_w, s_v) \in R$, and undefined
otherwise.
The $\mathsf{morph}$ constructor, and the \lift rule are generalized
to receive the initial index $x$, and postulate the existence of
an ending index $y$ in the postcondition, as follows.
\begin{mathpar}
\inferrule*[right=\liftx]
  {e : \spec{P}~A~\spec{Q}@V}
  {\mathsf{morph}\ f\ x\ e : \spec{\smorph {(f\ x)} P \bwedge I\ x}~A~\spec{\exists y\ldot \smorph {(f\ y)} Q \bwedge I\ y}@W}
\end{mathpar}


As an illustration, consider a \emph{history-based} specification of a
concurrent stack's $\mathsf{push}$ method~\cite{Sergey-al:ESOP15}.
%
\[
\begin{array}{c}
\act{push}(v) : \specK{[\thehist]}\ldot 
  \!\!\!\begin{array}[t]{l} 
        \spec{\lambda s\ldot \theheap_s(s) = \act{empty} \wedge \thehist_s(s) = \act{empty} \wedge \tau \sqsubseteq \thehist_o(s)}\\
        \spec{\lambda s\ldot \theheap_s(s) = \act{empty} \wedge \exists t\ vs\ldot \thehist_s(s) = t \hmapsto (vs, v\,{::}\,vs) \wedge \forall t'\in \mathsf{dom}(\thehist)\ldot t' < t}@\Stack
 \end{array}\\
\end{array}
\]
The $\Stack$ states have the fields $s = ((\theheap_s, \thehist_s),
(\theheap_j, \alpha), (\theheap_o, \thehist_o))$, where $\theheap_s,
\theheap_j, \theheap_o \in \heap$ and $\thehist_s, \thehist_o \in
\hist$.
The heaps $\theheap_s, \theheap_o$ are used to allocate new cells
before pushing them onto the stack. The heap $\theheap_j$ stores the
stack's physical layout, and $\alpha$ is the abstract contents of the
stack. The full definition of $\Sof{\Stack}$ is not important for the
discussion here; it suffices to know that we have a predicate
$\islist$ such that $\forall s\in\Sof{\Stack}\ldot
\islist{\alpha(s)}{\theheap_j(s)}$, i.e., $\islist$ describes how
$\alpha$ is laid out in $\theheap_j$.
%
%
Histories $\thehist_s$ and $\thehist_o$ are finite maps sending a
time-stamp $t$ to an abstract description of an operation performed at
time $t$. For example, the singleton history $42 \hmapsto (vs, \cons v
{vs})$, denotes that at time $42$, the element $v$ was pushed onto the
stack, thus changing $\alpha$ from the sequence $vs$ to $\cons v
{vs}$. Histories are a PCM under the operation of disjoint union
(undefined if operands share a time-stamp), and with the
$\mathsf{empty}$ history as unit.
%
If $t \in \mathsf{dom}(\thehist_s)$ (resp.~$t\in
\mathsf{dom}(\thehist_o)$), then the operation at time $t$ was
executed by the specified thread (resp.~the environment).
For example, $\mathsf{push}$ starts with $\thehist_s(s) =
\mathsf{empty}$ and ends with $\thehist_s(s) = t \hmapsto (vs, v::vs)$
to indicate that $\mathsf{push}(v)$ indeed pushed $v$. The interfering
threads may have executed their own operations before and after $t$,
to change the value of $\thehist_o$.  The conjunct $\forall t' \in
\mathsf{dom}(\thehist)\ldot t' < t$ temporally orders $t$ after the
timestamps of all the operations that terminated before
$\mathsf{push}(v)$ was invoked.


Now consider the program $e = \mathsf{push}(1) \parallel
\mathsf{push}(2)$, whose type derivation is in the Coq files. 
\[
e : \!\!\!\begin{array}[t]{l}
   \spec{\lambda s\ldot \theheap_s(s) = \mathsf{empty} \wedge \thehist_s(s) = \mathsf{empty}}\\ 
   \spec{\lambda s\ldot \theheap_s(s) = \mathsf{empty} \wedge \exists t_1\ {vs_1}\ t_2\ {vs_2}\ldot 
               \thehist_s(s) = t_1 \hmapsto (vs_1, 1\,{::}\,{vs_1}) \join 
                               t_2 \hmapsto (vs_2, 2\,{::}\,{vs_2})}@\Stack.
   \end{array}
\]
The specification reflects that $e$ pushes $1$ and $2$, to change the
stack contents from $vs_1$ to $1\,{::}\,{vs_1}$ at time $t_1$, and
from $vs_2$ to $2\,{::}\,{vs_2}$ at time $t_2$. The order of pushes is
unspecified, so we do not know if $t_1 < t_2$ or $t_2 < t_1$ (as
$\join$ is commutative, the order of $t_1$ and $t_2$ in the binding to
$\thehist_s(s)$ in the postcondition does not imply an ordering
between $t_1$ and $t_2$).  Moreover, we do not know that $t_1$ and
$t_2$ occurred in immediate succession (i.e., $t_2 = t_1 + 1 \vee t_1
= t_2 + 1$), as threads concurrent with $e$ could have executed
between $t_1$ and $t_2$, changing the stack arbitrarily. Thus, we also
cannot infer that the ending state of $t_1$ equals the beginning state
of $t_2$, or vice versa.


%

But what if we knew that $e$ is invoked \emph{quiescently}, i.e.,
without interfering threads? For example, a program working over the
resource $\Priv$ from Section~\ref{sec:overview} (hence, containing
only $\theheap_s$ and $\theheap_o$), can invoke $e$ over the empty
stack installed in $\theheap_s$. Because the stack is installed
privately, no threads other than the two children of $e$ can race on
it. Could we exploit quiescence, and derive \emph{just out of the
  specification} of $e$ that the stack at the end stores either the
list $[1, 2]$, or $[2, 1]$? The latter can even be stated without
histories, using solely heaps in $\Priv$, as follows.
\[
\spec{\lambda s\ldot \islist {\mathsf{nil}} {\theheap_s(s)}}
\ \spec{\lambda s\ldot \islist {[1, 2]} {\theheap_s(s)} \vee
  \islist {[2, 1]} {\theheap_s(s)}}@\Priv
\]

We would thus like a morphism $f : \Stack \rightarrow \Priv$ that
``erases histories'', but such a morphism cannot be constructed. Its
state component should map a $\Priv$-state, containing only heaps, to
a $\Stack$-state, containing histories as well, and thus has to
``invent'' the history component out of thin air.  This is where
families come in. We make $f : \Stack \rightarrowX{\hist} \Priv$ a
family over $X = \hist$, thereby passing to $f$ the history $\thehist$
that should be added to a $\Priv$ state in order to produce a $\Stack$
state. We define $f$'s state component as follows, where we use the
notation $(s_w, s_v) \in f\ \thehist$ instead of $f\ \thehist\ s_w = s_v$, to
emphasize the partiality of $f$.
\[
\begin{array}{rcl}
(s_\Priv, s_\Stack) \in f~\thehist & \eqdef & \theheap_s(s_\Priv) =
  \theheap_s(s_\Stack) \join \theheap_j(s_\Stack) \wedge
  \theheap_o(s_\Priv) = \theheap_o(s_\Stack) \wedge \hbox{}\\ & &
  \thehist_s(s_\Stack) = \thehist \wedge \thehist_o(s_\Stack) =
  \mathsf{empty}
\end{array}
\]
%
The first conjunct directly states that $\Stack$ is installed in
$\theheap_s(s_\Priv)$ by making one chunk of $\theheap_s(s_\Priv)$ be
the joint heap $\theheap_j(s_\Stack)$, and the other chunk be
$\theheap_s(s_\Stack)$.\footnote{As we want to build $s_\Stack$ out of
  $s_\Priv$, we have to identify a chunk of $\theheap_s(s_\Priv)$,
  which we want to assign to $\theheap_j(s_\Priv)$. Moreover, this
  chunk has to be unique, else $f$ will not satisfy the functionality
  property (2) of Definition~\ref{def:morph}. We ensure uniqueness by
  insisting that the predicate $\islist$ is precise -- a property
  commonly required in separation logics.}
The second conjunct says that the heap $\theheap_o(s_\Priv)$ of the
interfering threads is propagated to $\theheap_o(s_\Stack)$. The third
conjunct captures that the history component of $s_\Stack$ is set to
the index $\thehist$, as discussed immediately above. Finally, in the
last conjunct, the $\thehist_o(\Stack)$ history is declared
$\mathsf{empty}$, thus directly formalizing quiescence. We elide the
definition of $f$'s transition component, because we also elided the
definition of $\Stack$.\footnote{In our Coq files, we carried out the
  development for a Treiber variant of concurrent stacks, with some
  minor Treiber-specific modifications. We have also applied a similar
  morphism to a program constructing a spanning tree of a graph in
  place by marking and pruning the graphs' edges. There, the morphism
  was essential for showing that the tree constructed by pruning is
  spanning, i.e., it contains all the graph's nodes.}
%
%
Now, applying the \liftx rule to the $\Stack$ specification of $e$,
with $I~x$ being the always-true predicate on $\Priv$ states, and $x =
\mathsf{empty}$, gives us exactly the desired $\Priv$ specification,
after some trivial rearrangements.

\newcommand{\TaDA}{\textsc{TaDA}\xspace}
\newcommand{\Iris}{\textsc{Iris}\xspace}
\newcommand{\Disel}{\textsc{Disel}\xspace}
\newcommand{\CAP}{\textsc{CAP}\xspace}

\section{Related work}\label{sec:related}

There have been several approaches to relating concurrent resources,
including simultaneous modifications to their states, and program
lifting.

\paragraph{Higher-order auxiliary code.}
One approach, originated by Jacobs and
Piessens~\cite{Jacobs-Piessens:POPL11}, and later expanded by Svendsen
et al.~\cite{Svendsen-al:ESOP13, Svendsen-Birkedal:ESOP14}, relies on
parametrizing a program and its proof with auxiliary code that works
over the state of other resources.
For example, using the names from
Sections~\ref{sec:intro}~and~\ref{sec:overview}, a locking program
over $\Spin$ can be parametrized by an auxiliary function over $\Xfer$
which, once executed, transfers the shared heap in $\Xfer$ to private
state, much like the transition $\Xfer.\mathsf{open\_tr}$ would in
Section~\ref{sec:overview}. The locking program should be implemented
so as to invoke this auxiliary function at the moment of successful
locking.
In contrast, we formalized the scenario in Section~\ref{sec:overview}
by exhibiting a morphism from $\Spin$ to the extended resource $\CSL$
that couples $\Spin$ with $\Xfer$. Once $\Spin$ locks, the heap
transfer in $\CSL$ does not occur automatically, but the $\CSL$
resource is placed in a state where the transfer can be executed by
invoking $\mathsf{open\_tr}$. This is somewhat less immediate than
parametrization, but sufficient for our main goal, which is reusing
$\Spin$'s implementation of locking without reverification.
One advantage of our approach is that lifting a program from the
source to the target resource is done after the program has been
implemented, and only depends on the program's type (i.e., the
pre/postcondition, and the definition of the two resources), whereas
with parametrization, the program has to be developed with the
parameter auxiliary functions in mind from the very beginning. A
well-known challenge of parametrizing a program by an auxiliary
function is exhibited when the point at which to execute the auxiliary
function can be determined only after the program has already
terminated.
%
%
We \emph{expect} our morphisms to scale to such cases, precisely
because lifting depends only on the program's type, not the code
(hence, termination is irrelevant). However, this remains to be
confirmed.





\paragraph{Abstract atomicity.}
Another approach, originated by Da Rocha-Pinto et
al.~\cite{ArrozPincho-al:ECOOP14} in \TaDA logic, and recently adopted
by \Iris~\cite{Jung-al:POPL15}, introduces a new judgment form,
$\langle P\rangle\ e\ \langle Q\rangle$, capturing that $e$ has a
precondition $P$ and postcondition $Q$, but is also \emph{abstractly
  atomic} in the following sense: $e$ and its concurrent environment
maintain the validity of $P$ through the execution, until at one point
$e$ makes an atomic step that makes $Q$ hold. After that point, $Q$
may be invalidated, either by future steps of $e$, or by the
environment. The challenge of this approach is that the new judgment
has a rather complicated proof theory, and
%
comes with auxiliary concepts, such as atomicity tokens, that impose
some restrictions. 
%
%
For example, programs with helping, where one thread executes the work
on behalf of another, currently are not supported by \TaDA because
their verification requires atomicity tokens to exchange ownership.
In contrast, for us, ownership transfer is encoded by transition
coupling, and is thus directly addressed by morphisms and
simulations. We have been able to easily support helping, and have
verified, in our Coq files, the flat combiner
algorithm~\cite{Hendler-al:SPAA10}, a non-trivial helping example.  We
also verified representative clients that couple the transitions of
the flat combiner with non-idle transitions of another resource. These
latter transitions are to be executed simultaneously with the
flat-combiner helping.
%
%
The abstract atomicity approach, either in \TaDA or \Iris, also does
not consider simulation as a way of relating resources.

The \Iris version of abstract atomicity differs from the one of \TaDA
in that it is encoded using higher-order state available in \Iris's
model. Otherwise, the fragment of \Iris's proof theory that handles
abstract atomicity is almost identical to that of \TaDA. Similarly to
\SCSL~\cite{LeyWild-Nanevski:POPL13}, \FCSL~\cite{Nanevski-al:ESOP14},
and the current paper, \Iris uses PCMs to encode auxiliary
state. \Iris also encodes STSs via PCMs, but that is a move that we
resist here. The structure-preserving functions between PCMs
(aka.~\emph{local actions}~\cite{Calcagno-al:LICS07}) are
significantly different from structure-preserving functions between
STSs that we consider here in the form of morphisms, which is why we
avoid conflating the two. Finally, while in this paper we do not
consider higher-order state, we expect that our morphism-based
approach should easily reconcile with it. In particular, we expect
that the $\lift$ rule could be proved sound in \Iris's model (if
extended with morphisms), but this is an orthogonal consideration.

\paragraph{Protocol hooks.} 
Concurrently with us, Sergey et al.~\cite{ser+al:oopsla17} have
designed a logic \Disel for distributed systems, in which one can
combine distributed protocols---represented as STSs---by means of
\emph{hooks}. A hook on a transition $t$ prevents $t$ from execution,
unless the condition associated with the hook is satisfied. In this
sense, hooks implement a form of our transition coupling, but where
one operand is the idle transition $\mathsf{id\_tr}~P$, with $P$ the
associated condition. The above version of \Disel does not consider
transition coupling where both operands are non-idle (which we needed
in Figure~\ref{fig:rwlock} to define, for example, the
$\mathsf{toreader\_tr}$ transition, and in the flat combiner
implementation in our Coq files), or notions of morphism and
simulation. Our work does not consider distributed protocols.


\paragraph{Refinement reasoning and linearizability.}
In a somewhat different, relational, flavor of separation
logics~\cite{Liang-al:POPL12,Liang-Feng:PLDI13,Turon-al:ICFP13}, and
more generally, in the work on proving
linearizability~\cite{HenzingerSV+CONCUR13,SchellhornWD+CAV12,BouajjaniEEM+cav17},
the approaches explicitly establish a simulation between two programs;
typically one concurrent, the other sequential. This is required for
showing that a concurrent program is logically atomic; that is, it
linearizes to the given sequential program.
%
%
%
Our goal in this paper is somewhat different. Instead of establishing
a simulation between two programs, we establish a simulation (i.e., a
morphism) between two STSs, which are components of program types,
but are themselves not programs.
Simulation between STSs is easier to establish than simulation between
programs, as STSs have a much simpler structure---being transition
systems, they omit programming constructions such as conditionals,
loops, local state, or function calls.
Thus, our simulation does not \emph{directly} prove that a program is
linearizable, but is intended for lifting a program from the source to
the target STS, without reproving. Logical atomicity should be handled
by other components of the system. For example, recent related work on
\FCSL~\cite{del+ser+nan+ban:ecoop17}, shows that specifications based
on PCMs with \emph{self} and \emph{other} components can specify
logical atomicity, even for sophisticated algorithms with
future-dependent linearization points~\cite{Jayanti:STOC05}.

\paragraph{Previous work on \FCSL}
The current paper builds on the previous work on
\FCSL~\cite{Nanevski-al:ESOP14}, to which it adds a novel notion of
morphism, and significantly modifies the definition of concurrent
resources. In \FCSL, each concurrent resource was a finite map from
labels (natural numbers) to sub-components. For example, using the
concepts from Section~\ref{sec:overview}, one could represent $\CSL$
as a finite map $l_1 \hmapsto \Spin~{\uplus}~l_2 \hmapsto \Xfer$,
where $l_1$ and $l_2$ are labels identifying $\Spin$ and $\Xfer$,
respectively. This approach provides interesting equations on
resources; for example, one can freely rearrange the finite map
components by using commutativity and associativity of
$\dotcup$. However, it also complicates mechanized verification,
because one frequently needs to prove that a label is in the domain of
a map, before extracting the labeled component. In the new version of
\FCSL, we significantly reduce the sizes of mechanized proofs by
removing labels and combining components by means of pairing their
states (Definitions~\ref{def:stprod} and~\ref{def:products} in
Section~\ref{sec:formal}). Consequently, if we changed the definition
of $\CSL$ in Section~\ref{sec:overview} into $\CSL'$ by commuting
$\Spin$ and $\Xfer$ throughout the construction, then $\CSL$ and
$\CSL'$ would not be \emph{equal} resources, but they will be
\emph{isomorphic}, in that we could exhibit cancelling morphisms
between the two. But this requires first having a notion of morphism,
which is one of the technical contributions of this paper. Previously,
\FCSL supported quiescence by means of a dedicated and complex
inference rule.
%
%
In Section~\ref{sec:param}, we show that quiescence reduces to $\liftx$
rule, via indexed morphism families.


\section{Conclusions and future work}\label{sec:conclusions}

This paper argues that a notion of simulation to relate
resources, and the corresponding notion of morphism that allows
lifting programs, are key components of modular reasoning about
concurrent programs. We apply these notions in \FCSL, a separation
logic for fine-grained concurrency. Our preliminary experiments
indicate that the formalism leads to significant shortening of
mechanized proofs and reuse of resource definitions and program
verifications.
%
%
Given a morphism from resource $V$ to resource $W$, programs written
over $V$ can automatically be lifted to work over $W$, and the lifting
is realized by means of a single Hoare-style inference rule.  We call
our notion of morphism ``subjective simulation'', because it applies
to STSs with subjective division of states into \emph{self} and
\emph{other} components. 
A morphism exhibits a form of forward simulation~\cite{aba+lam:91} of
$V$ by $W$. The morphism is also interference-aware, as it exhibits a
form of simulation of $W$ by $V$, performed on transposed states,
where the \emph{self} and \emph{other} components are swapped.

Morphisms are useful for a number of applications. One is lifting a
program from $V$ to $W$, when $W$ includes $V$ as a
sub-component. This was illustrated in Section~\ref{sec:rwlock}, where
we built a resource for readers/writers lock in a staged, decomposed,
manner.
Another application is in managing the scope of auxiliary state. This
was illustrated in Section~\ref{sec:param}, where auxiliary state of
histories is introduced within the scope of a morphism that maps
abstract stacks to their underlying heaps. Such histories should be
invisible to the clients, which should only view the underlying
modifications to the private heaps. This application required a
generalization to indexed morphism families, and could also encode
quiescence.
%
In the Coq files, we have further verified a flat combiner and an
in-place construction of a spanning tree of a graph.
%

Beyond the progress reported here, we expect that our notion of morphisms will have many other applications as well. In the immediate future, we plan to apply
morphisms to procedures with linearization points whose placement in
time can be determined only after the procedure's
termination~\cite{Jayanti:STOC05,del+ser+nan+ban:ecoop17}. Most
related work deals with such programs by formalizing the dependence of
the linearization points on the future events as a form of
non-determinism, and the corresponding proofs employ features such as
prophecy variables~\cite{aba+lam:91} (equivalently, speculations,
backward simulations), which have not been reconciled with program
lifting. It has recently been
argued~\cite{del+ser+nan+ban:ecoop17,BouajjaniEEM+cav17} that
future-dependence may not need non-determinism, as the placement of
the linearization points can be deterministically resolved at the
level of proofs. Thus, we expect that morphisms and \FCSL will
directly apply.
%


\bibliographystyle{plain}
\bibliography{bibmacros,references,proceedings}

\appendix
\section{Generalized definitions for indexed morphism families}\label{sec:famdef}
In this appendix, we show how the definitions of morphism,
$f$-stepping and $f$-stability, generalize to indexed families. When
$X$ is the unit type, we recover the morphism-related definitions from
Section~\ref{sec:formal}.
%

\begin{definition}[Indexed family of morphisms]\label{def:pmorph}
An indexed family of morphisms $f : V \rightarrowX X W$ (or just
family), consists of two components:
\begin{itemize}
\item A function from $x \in X$ to relation on the states of $V$ and
  $W$, which we write as $(s_v, s_w) \in f\ x$, where $s_v$ is a
  $V$-state, and $s_w$ is a $W$-state.
\item A function mapping $x \in X$ and an internal transition of $V$
  to internal transitions of $W$, which we write as $f\ x: \intof{V}
  \rightarrow \intof{W}$.
\end{itemize}
The components satisfy the following properties:
\begin{enumerate}
\item\label{pmorph:simWV} \emph{($W$ simulates $V$ by internal steps)}
  if $t \in \intof{V}$ and $t\ s_v\ s'_v$ and $(s_v, s_w) \in f\ x$, then
  there exists $x'$, $s'_w$ such that $f\ x\ t\ s_w\ s'_w$ and $(s'_v, s'_w)
  \in f\ x'$.
\item\label{pmorph:simVW} \emph{($V$ simulates $W$ by other steps)} if
  $s_w \osteps[W]{} s'_w$ and $(s_v, s_w) \in f\ x$, then there exists
  $s'_v$ such that $s_v \osteps[V]{} s'_v$ and $(s'_v, s'_w) \in f\ x$.
\item\label{pmorph:contra} \emph{(functionality)}: if $(s_{v1}, s_w)
  \in f\ x$ and $(s_{v2}, s_w) \in f\ x$, then $s_{v1} = s_{v2}$
\item\label{pmorph:frame} \emph{(frame preservation)} there exists
  function $\phi : U_W \rightarrow U_v$ (notice the contravariance),
  such that: if $(s_v, s_w \zig p) \in f\ x$, then $s_v = s'_v \zig
  (\phi\ p)$ for some $s'$, and $(s'_v \zag \phi\ p, s_w \zag p) \in
  f\ x$.
\item\label{pmorph:other} \emph{(other-fixity)} if $(s_v, s_w) \in f\ x$
  and $(s'_v, s'_w) \in f\ x'$ and $a_o(s_w) = a_o(s'_w)$ then $a_o(s_v) =
  a_o(s'_v)$.
\item\label{pmorph:backw} \emph{(index injectivity)} if $(s_v, s_{w1})
  \in f\ x_1$ and $(s_v, s_{w2}) \in f\ x_2$ then $x_1 = x_2$
\end{enumerate}
\end{definition}

In most of the properties of Definition~\ref{def:pmorph}, the index
$x$ is propagated unchanged. The only properties where $x$ is
significant are (\ref{pmorph:simWV}) and the new property
(\ref{pmorph:backw}). Compared to Definition~\ref{def:morph}, the
property (\ref{pmorph:simWV}) allows that $x$ changes into $x'$ by a
transition. In the $\Stack$ example in Section~\ref{sec:param}, if we
lift $e$ by using the index $x = \mathsf{empty}$ (i.e., write
$\mathsf{morph}\ \mathsf{empty}\ f\ e$), then this index will evolve
with $e$ taking the transitions of $\Stack$ to track how $e$ changes
the self history by adding the entries for pushing 1 and 2.  The
property (\ref{pmorph:backw}) requires that $s_v$ uniquely determines
the index $x$. In the $\Stack$ example, it is easy to see that the
definition of $f$ satisfies this property, because equal states have
equal histories.


\begin{definition}[$f$-stepping]
Let $f : V \rightarrowX X W$ be a family, and let $s_w$, $s'_w$ be
$W$-states. We say that $x, s_w$ $f$-steps to $x', s'_w$, written $x,
s \mstep[f]{} x', s'$, if one of the following is true:
\begin{enumerate}
\item exists $t \in \intof{V}$ and $s_v$, $s'_v$, such that $(s_v, s_w) \in f\ x$, $(s'_v, s'_w) \in f\ x'$, $t\ s_v\ s'_v$ and $f\ x\ t\ s_w\ s'_w$
\item $s_w \ostep[W]{} s'_w$
\end{enumerate}
In other words, $x, s_w$ steps by $f$ into $x', s'_w$, either if it
steps by ordinary interference on $W$, or the step is an $f$-image of
a step by an internal transition in $V$. We write $\msteps[f]{}$ for
reflexive-transitive closure of $\mstep[f]{}$.
\end{definition}

\begin{definition}[$f$-stability]
Let $f : V \rightarrowX X W$ be a family. A predicate $P$ over $X$ and
$W$-states is \emph{$f$-stable in state $x, s$} if whenever $x, s
\msteps[f]{} x', s'$, then $P\ x'\ s'$. Predicate $P$ is $f$-stable if
it is $f$-stable in state $x, s$ for every $x, s$ for which $P\ x\ s$.
Given a predicate $P$ over $X$ and $W$-states, we define its
\emph{$f$-stabilization} $\fstab{P}{f}$ as the following predicate:
\[
\fstab{P}{f}\ x\ s \eqdef \forall x'\ s'\ldot x, s \msteps[f]{} x', s' \rightarrow P\ x'\ s'.
\]
\end{definition}

\section{Denotational semantics}\label{sec:model}

\newcommand{\ty}{~{:}~}
\newcommand{\Ghi}{\widehat{\Phi}}
\newcommand{\entangle}{\rtimes}
\newcommand{\tsteps}[1]{~{\rightarrowX{#1}}~}
\newcommand{\Path}{\pi}
\newcommand{\Paths}{\zeta}
\newcommand{\ChoiceAct}{\mathsf{ChoiceAct}}
\newcommand{\SeqRet}{\mathsf{SeqRet}}
\newcommand{\SeqStep}{\mathsf{SeqStep}}
\newcommand{\ParRet}{\mathsf{ParRet}}
\newcommand{\ParL}{\mathsf{ParL}}
\newcommand{\ParR}{\mathsf{ParR}}
\newcommand{\MorphStep}{\mathsf{MorphStep}}
\newcommand{\MorphRet}{\mathsf{MorphRet}}
\newcommand{\HideStep}{\mathsf{HideStep}}
\newcommand{\HideRet}{\mathsf{HideRet}}
\newcommand{\InjRet}{\mathsf{InjRet}}
\newcommand{\InjStep}{\mathsf{InjStep}}
\newcommand{\SCST}{our system}

\newcommand{\rely}[1]{\mathcal{R}_{#1}}
\newcommand{\guar}[1]{\mathcal{G}_{#1}}

\newcommand{\unwind}{\mathsf{unwind}}
\newcommand{\pathtp}{\mathsf{path}}
\newcommand{\tsafe}{\mathsf{safe}}

\newcommand{\set}[1]{\left\{{#1}\right\}}

Our semantic model largely relies on the denotational semantic of
\emph{action trees}~\cite{LeyWild-Nanevski:POPL13}. A tree implements
a finite partial approximation of program behavior; thus a program of
type $\mathsf{ST}\ V\ A$ will be denoted by a set of such trees. The
set may be infinite, as some behaviors may only be reached in the
limit, after infinitely many finite approximations.

An action tree is a generalization of the Brookes' notion of action
trace in the following sense. Where action trace semantics approximate
a program by a set of traces, we approximate with a set of trees. A
tree differs from a trace in that a trace is a sequence of actions and
their results, whereas a tree contains an action followed by a
\emph{continuation} which itself is a tree parametrized wrt.~the
output of the action.

In this appendix, we first define the denotation of each of our
commands as a set of trees. Then we define the semantic behavior for
trees wrt.~resource states, in a form of operational semantics for
trees. Then we relate this low-level operational semantics of trees to
high-level transitions of a resource by an $\mathsf{always}$ predicate
(Section~\ref{sec:modal-pred}) that ensures that a tree is resilient
to any amount of interference, and that all the operational steps by a
tree are \emph{safe}.
The $\mathsf{always}$ predicate will be instrumental in defining the
$\vrf$-predicate transformer from Section~\ref{sec:formal}, and from
there, in defining the type of Hoare triples
$\spec{P}\ A\ \spec{Q}@V$.
Both the $\mathsf{ST}\ V\ A$ type and the Hoare triple type will be
\emph{complete lattices} of sets of trees, giving us a suitable
setting for modeling recursion.
The \emph{soundness} of \FCSL follows from showing that the lemmas
about the $\vrf$ predicate transformer listed in
Section~\ref{sec:formal}, are satisfied by the denotations of the
commands.


We choose the Calculus of Inductive Constructions
(CiC)~\cite{coq-team,Bertot-Casteran:04} as our meta logic.  This has
several important benefits.  First, we can define a \emph{shallow
  embedding} of our system into CiC that allows us to program and
prove directly \emph{with the semantic objects}, thus immediately
lifting to a full-blown programming language and verification system
with higher-order functions, abstract types, abstract predicates, and
a module system.  We also gain a powerful dependently-typed
$\lambda$-calculus, which we use to formalize all semantic definitions
and meta theory, including the definition of action trees by
\emph{iterated inductive definitions}~\cite{coq-team},
specification-level functions, and programming-level higher-order
procedures.
Finally, we were able to mechanize the entire semantics and meta theory
in the Coq proof assistant implementation of~CiC.

\subsection*{Action trees and program denotations}
\label{sec:opsem}
\begin{definition}[Action trees]\label{def:trees}
The type $\mathsf{tree}~V~A$ of $A$-returning action trees is defined
by the following iterated inductive definition.
\[
\begin{array}{rcl}
\mathsf{tree}~V~A & \eqdef & \mathsf{Unfinished}\\
& \mid & \mathsf{Ret}\ (v \ty A)\\
& \mid & \mathsf{Act}~(a : \mathsf{action}~V~A)\\
& \mid & \mathsf{Seq}~(T \ty \mathsf{tree}~V~B)~(K \ty B \rightarrow \mathsf{tree}~V~A)\\
& \mid & \mathsf{Par}~(T_1 \ty \mathsf{tree}~V~B_1)\ (T_2 \ty \mathsf{tree}~V~B_2)\ (K \ty B_1 \times B_2 \rightarrow \mathsf{tree}~V~A)\\
& \mid & \mathsf{Morph}~(x : X)~(f : W \rightarrowX X V)~(T \ty \mathsf{tree}~W~A)
\end{array}
\]  
\end{definition}

Most of the constructors in Definition~\ref{def:trees} are
self-explanatory.
Since trees have finite depth, they can only approximate potentially
infinite computations, thus the $\mathsf{Unfinished}$ tree indicates
an incomplete approximation. %
$\mathsf{Ret}\ v$ is a terminal computation that returns value
$v\,{:}\,A$. %
The constructor $\mathsf{Act}$ takes as a parameter an action $a :
\mathsf{action}~V~A$, as defined in Section~\ref{sec:formal}.
$ \mathsf{Seq}~T~K$ sequentially composes a $B$-returning tree $T$
with a continuation $K$ that takes $T$'s return value and generates
the rest of the approximation.
$\mathsf{Par}\ T_1\ T_2\ K$ is the parallel composition of trees $T_1$
and $T_2$, and a continuation $K$ that takes the pair of their results
when they join.  CiC's iterated inductive definition permits the
recursive occurrences of $\mathsf{tree}$ to be \emph{nonuniform}
(e.g., $\mathsf{tree}\ B_i$ in $\mathsf{Par}$) and \emph{nested}
(e.g., the \emph{positive} occurrence of $\mathsf{tree}\ A$ in the
continuation). Since the CiC function space includes case-analysis,
the continuation may branch upon the argument.
%
The $\mathsf{Morph}$ constructor embeds an index $x:X$, morphism $f :
W \rightarrowX X V$, and tree $T \ty \mathsf{tree}~W~A$ for the
underlying computation. The constructor will denote $T$ should be
executed so that each of its actions is modified by $f$ with an index
$x$.
We can now define the denotational model of our programs; that is the
type $\mathsf{ST}~V~A$ of sets of trees, containing
$\mathsf{Unfinished}$.
\[
\mathsf{ST}~V~A \eqdef \{e : \mathsf{set}~(\mathsf{tree}~V~A) \mid \mathsf{Unfinished} \in e\}
\]

The denotations of the various constructors combine the trees of the
individual denotations, as shown below.
\[
\begin{array}{r@{\ }c@{\ }l}
\mathsf{ret}\ (r \ty A) & \eqdef & \{\mathsf{Unfinished}, \mathsf{Ret}~r\}
\\
x \leftarrow e_1; e_2 & \eqdef & \{\mathsf{Unfinished}\} \cup 
%
%
\{\mathsf{Seq}~T_1~K \mid T_1 \in e_1 \wedge \forall x \ldot K~x \in e_2\}
\\
e_1 \parallel e_2 & \eqdef & \{\mathsf{Unfinished}\} \cup 
\{\mathsf{Par}~T_1~T_2~\mathsf{Ret} \mid T_1 \in e_1 \wedge T_2 \in e_2\}
\\
\mathsf{atomic}~a & \eqdef & \{\mathsf{Unfinished}, \mathsf{Act}~a\}
\\
\mathsf{morph}~x~f~e & \eqdef & \{\mathsf{Unfinished}\} \cup 
\{\mathsf{Morph}~x~f~T \mid T \in e\}
\end{array}
\]
The denotation of $\mathsf{ret}$ simply contains the trivial
$\mathsf{Ret}$ tree, in addition to $\mathsf{Unfinished}$, and
similarly in the case of $\mathsf{act}$. The trees for sequential
composition of $e_1$ and $e_2$ are obtained by pairing up the trees
from $e_1$ with those from $e_2$ using the $\mathsf{Seq}$ constructor,
and similarly for parallel composition and morphism application.

The denotations of composed programs motivate why we denote programs
by non-empty sets, i.e., why each denotation contains at least
$\mathsf{Unfinished}$. If we had a program $\mathsf{Empty}$ whose
denotation is the empty set, then the denotation of $x \leftarrow
\mathsf{Empty}; e'$, $\mathsf{Empty} \parallel e'$ and
$\mathsf{morph}~x~f~\mathsf{Empty}$ will all also be empty, thus
ignoring that the composed programs exhibit more behaviors. For
example, the parallel composition $\mathsf{Empty} \parallel e'$ should
be able to evaluate the right component $e'$, despite the left
component having no behaviors.

By including $\mathsf{Unfinished}$ in all the denotations, we ensure
that behaviors of the components are preserved in the composition. For
example, the parallel composition $\{\mathsf{Unfinished}\} \parallel
e'$ is denoted by the set below which contains an image of each tree
from $e'$, thus capturing the behaviors of $e'$.
\[\{\mathsf{Unfinished}\} \cup
\{\mathsf{Par}~\mathsf{Unfinished}~T~\mathsf{Ret} \mid T \in e'\}\]

\subsection*{Operational semantics of action trees}

The judgment for small-step operational semantics of action trees has
the form $\Delta \vdash \bar{x}, s, T \tsteps{\Path} \bar{x}', s', T'$
(Figure~\ref{fig:tsteps}). We explain the components of this judgment
next.

First, the component $\Delta$ is a morphism context. This is a
sequence, potentially empty, of morphism families
\[
f_0 : V_1 \rightarrowX {X_0} W, f_1 : V_2 \rightarrowX {X_1} V_1, \ldots, f_n : V \rightarrowX {X_n} V_n
\]
We say that $\Delta$ has resource type $V \rightarrow W$, and index
type $(X_0, \cdots, X_n)$. An empty context $\cdot$ has resource type
$V \rightarrow V$ for any $V$.

Second, the components $\bar{x}$ and $\bar{x}'$ are tuples, of type
$(X_0, \cdots, X_n)$, and we refer to them as indexes. Intuitively,
the morphism context records the morphisms under which a program
operates. For example, if we wrote a program of the form
\[\mathsf{morph}\ f_0\ x_0\ (\cdots (\mathsf{morph}\ f_n\ x_n\ e) \cdots),\]
it will be that the trees that comprise $e$ execute under the morphism
context $f_0, \ldots, f_n$, with an index tuple $(x_0, \ldots, x_n)$.

Third, the components $s$ and $s'$ are $W$-states, and $T, T' :
\mathsf{tree}~V~A$, for some $A$. The meaning of the judgment is that
a tree $T$, when executed in a state $s$, under the context of
morphisms $\Delta$ produces a new state $s'$ and residual tree $T'$,
encoding what is left to execute. The resource of the trees and the
states disagree (the states use resource $W$, the trees use $V$), but
the morphism context $\Delta$ relates them as follows. Whenever the
head constructor of the tree is an action, the action will first be
morphed by applying all the morphisms in $\Delta$ in order, to the
transitions that constitute the head action, supplying along the way
the projections out of $x$ to the morphisms. This will produce a new
index $x'$ and an action on $W$-states, which can be applied to $s$ to
obtain $s'$.

Fourth, the component $\Path$ is of $\pathtp$ type, identifying the
position in the tree where we want to make a reduction.
\[
\begin{array}{rclclclcl}
\pathtp & \eqdef & \ChoiceAct &~~|~~& \SeqRet &~~|~~& \SeqStep~(\Path : \pathtp) &~~|~~&   \\
&&               \ParRet &~~|~~& \ParL~(\Path : \pathtp) &~~|~~& \ParR~(\Path : \pathtp) &~~|~~& \\
&&               \MorphRet &~~|~~& \MorphStep~(\Path : \pathtp).
\end{array}
\label{eq:path}
\]
The key are the constructors $\ParL~\Path$ and $\ParR~\Path$. In a
tree which is a $\mathsf{Par}$ tree, these constructors identify that
we want to reduce in the left and right subtree, respectively,
iteratively following the path $\Path$. If the tree is not a
$\mathsf{Par}$ tree, then $\ParL$ and $\ParR$ constructors will not
form a good path; we define further below when a path is good for a
tree. The other path constructors identify positions in other kinds of
trees. For example, $\ChoiceAct$ identifies the head position in the
tree of the form $\mathsf{Act}(a)$, $\SeqRet$ identifies the head
position in the tree of the form $\mathsf{Seq}~(\mathsf{Ret}~v)~K$
(i.e., it identifies a position of a beta-reduction), $\SeqStep~\pi$
identifies a position in the tree $\mathsf{Seq}~T~K$, if $\pi$
identifies a position within $T$, etc. We do not paths for trees of
the form $\mathsf{Unfinished}$ and $\mathsf{Ret}~v$, because these do
not reduce.

In order to define the operational semantics on trees, we next require
a few auxiliary notions.  First, we need a function
$\Delta(\bar{x})(t)$ that morphs an internal transition $t$ of a
resource $V$, into a transition of a resource $W$, by iterating the
morphisms in the context $\Delta$ of resource type $V \rightarrow W$,
and passing along the elements out of the tuple $\bar{x}$ of type
$(X_0,\cdots, X_n)$. The function is defined by induction on the
structure of $\Delta$, as follows.
\[
\begin{array}{rcl}
(\cdot)~()~(t) & \eqdef & t\\
(f_0 : V_1 \rightarrowX {X_0} W, \Delta)~(x_0, \bar{x})~t & \eqdef & f_0~x_0~(\Delta~\bar{x}~t)
\end{array}
\]
That is, if $\Delta$ is the empty context, the index is empty tuple
$()$. In that case, there is nothing to do, so we just return the
transition $t$. Otherwise, we strip the first morphism $f_0$ from the
context, and the first index component $x_0$, iterate the construction
on the smaller context and index tuple, and apply $f_0\ x_0$ to the
result of the iterated construction.

Second, we need to have a similar iterative construction on states as
well, which will transforms the states according to morphisms in
$\Delta$.  We write $\mathsf{unwind}\ \Delta\ t\ x\ s\ x'\ s'$ to
denote that the the transition $t$ of the resource $V$ steps from the
$W$-state $s$ to $W$-state $s'$ in the morphism context $\Delta$. The
notion is again defined by induction on the structure of $\Delta$, as
follows:
\[
\begin{array}{rcl}
\unwind~\cdot~t~()~s~()~s' & \eqdef & t\ s\ s'\\
\unwind~(f_0 : V_1 \rightarrowX {X_0} W, \Delta)~t~(x_0, \bar{x})~s~(x'_0, \bar{x}')~s' & \eqdef &
\Delta~(x_0, \bar{x})~t~s~s' \wedge \exists s_1\ s'_1\ldot (s_1, s) \in f_0\ x_0 \wedge \hbox{}\\
& & \unwind~\Delta~{\bar{x}}~s_1~{\bar{x}'}~s'_1 \wedge (s'_1, s') \in f_0\ x'_0 
\end{array}
\]
If $\Delta$ is the empty context, there is nothing to do, and we just
return $t\ s\ s'$. Otherwise, we require that $s$ and $s'$ are related
by the image transition $\Delta~(x_0, \bar{x})~t$, but also that we
can iteratively produce image states of $s$ and $s'$ under all the
morphisms in the context.

We will frequently use the judgment in the case when $\Delta$ is the
empty context, and correspondingly, $\bar{x}$ and $\bar{x}'$ are empty
tuples $()$. In that case, we abbreviate, and write the judgment
simply as
\[
s, T \tsteps{\Path} s', T'.
\]

\begin{figure}[t] 
{\small
\begin{mathpar}
\inferrule
 {\unwind~\Delta~(a~v)~\bar{x}~s~\bar{x}'~s'}
 {\Delta \vdash \bar{x}, s, \mathsf{Act}~a \tsteps{\mathsf{\ChoiceAct}} \bar{x}', s'; \mathsf{Ret}~v}

\inferrule
 {~}
 {\Delta \vdash \bar{x}, s, \mathsf{Seq}~(\mathsf{Ret}~v)~K \tsteps{\SeqRet} \bar{x}, s, K~v}

\inferrule
 {\Delta \vdash \bar{x}, s, T \tsteps{\Path} \bar{x}, s', T'}
 {\Delta \vdash \bar{x}, s, \mathsf{Seq}~T~K \tsteps{\SeqStep~\Path} \bar{x}', s', \mathsf{Seq}~T'~K}

\inferrule
 {~}
 {\Delta \vdash \bar{x}, s, \mathsf{Par}\ (\mathsf{Ret}\ v_1)\ (\mathsf{Ret}\ v_2)\ K \tsteps{\ParRet}
  \Delta \vdash \bar{x}, s, K\ (v_1, v_2)}

\inferrule
 {\Delta \vdash \bar{x}, s, T_1 \tsteps{\Path} \bar{x}', s', T'_1}
 {\Delta \vdash \bar{x}, s, \mathsf{Par}\ T_1\ T_2\ K \tsteps{\ParL~\Path}
  \Delta \vdash \bar{x}', s', \mathsf{Par}\ T'_1\ T_2\ K}

\inferrule
 {\Delta \vdash \bar{x}, s, T_2 \tsteps{\Path} \bar{x}', s', T'_2}
 {\Delta \vdash \bar{x}, s, \mathsf{Par}\ T_1\ T_2\ K \tsteps{\ParR~\Path}
  \Delta \vdash \bar{x}', s', \mathsf{Par}\ T_1\ T'_2\ K}

\inferrule
 {~}
 {\Delta \vdash \bar{x}, s, \mathsf{Morph}~f~y~(\mathsf{Ret}~v) \tsteps{\MorphRet} \bar{x}, s, \mathsf{Ret}~v}

\inferrule
 {\Delta, f \vdash (\bar{x}, y), s, T \tsteps{\Path} (\bar{x}', y'), s', T'}
 {\Delta \vdash \bar{x}, s, \mathsf{Morph}~f~y~T \tsteps{\MorphStep~\Path} \bar{x}', s', \mathsf{Morph}~f~y'~T'}
\end{mathpar}}
 \caption{Judgment $\Delta \vdash \bar{x}, s, T \tsteps{\Path}
   \bar{x}', s', T'$, for operational semantics on trees, which reduces
   a tree with respect to the path $\Path$.}
 \label{fig:tsteps}
\end{figure} 

The operational semantics on trees in Figure~\ref{fig:tsteps} may not
make a step on a tree for two different reasons. The first, benign,
reason is that the the chosen path $\Path$ does not actually determine
an action or a redex in the tree $T$. For example, we may have $T =
\mathsf{Unfinished}$ and $\Path = \mathsf{ParR}$. But we can choose
the right side of a parallel composition only in a tree whose head
constructor is $\mathsf{Par}$, which is not the case with
$\mathsf{Unfinished}$.  We consider such paths that do not determine
an action or a redex in a tree to be ill-formed.
The second reason arises when $\Path$ is actually well-formed. In that
case, the constructors of the path uniquely determine a number of
rules of the operational semantics that should be applied to step the
tree. However, the premises of the rules may not be satisfies. For
example, in the $\mathsf{ChoiceAct}$ rule, there may not exist a $v$
such that $\unwind~\Delta~(a~v)~\bar{x}~s~\bar{x}'~s'$.
To differentiate between these two different reasons, we first define
the notion of well-formed, or \emph{good} path, for a given tree.
\begin{definition}[Good paths and safety]
\label{def:good}
Let $T \ty \mathsf{tree}~V~A$ and $\Path$ be a path.  Then the
predicate $\mathsf{good}~T~\Path$ is defined as follows:
\[
\begin{array}{rllcl}
  \mathsf{good} & (\mathsf{Act}~a) & \ChoiceAct & ~\eqdef~ & \mathsf{true}   
  \\
  \mathsf{good} & (\mathsf{Seq}~(\mathsf{Ret}~v)~\_) & \SeqRet & ~\eqdef~ & \mathsf{true}  
  \\
  \mathsf{good} & (\mathsf{Seq}~T~\_) & \SeqRet~\Path & ~\eqdef~ & \mathsf{good}~T~\Path 
  \\
  \mathsf{good} & (\mathsf{Par}~(\mathsf{Ret}~\_)~(\mathsf{Ret}~\_)~\_) & \ParRet & ~\eqdef~ & \mathsf{true}  
  \\
  \mathsf{good} & (\mathsf{Par}~T_1~T_2~\_) & \ParL~\Path & ~\eqdef~ & \mathsf{good}~T_1~\Path
  \\
  \mathsf{good} & (\mathsf{Par}~T_1~T_2~\_) & \ParR~\Path & ~\eqdef~ & \mathsf{good}~T_2~\Path
  \\
  \mathsf{good} & (\mathsf{Morph}~f~x~(\mathsf{Ret}~\_)) & \MorphRet & ~\eqdef~ & \mathsf{true}
  \\
  \mathsf{good} & (\mathsf{Morph}~f~x~T) & \MorphStep~\Path & ~\eqdef~ & \mathsf{good}~T~\Path \\
  \mathsf{good} & T & \Path & ~\eqdef~ & \mathsf{false} \text{~~otherwise}
\end{array}
\]
We now say that a state $s$ is safe for the tree $T$ and path $\Path$,
written $s \in \tsafe~t~\Path$ if:
\[
\mathsf{good}~T~\Path \rightarrow \exists s'\ T'\ldot s, T \tsteps{\Path} s', T'
\]
\end{definition}

Notice that in the above definition, the trees $\mathsf{Unfinished}$
and $\mathsf{Ret}\ v$ are safe for any path, simply because there are
no good paths for them, as such trees are terminal. On the other hand,
a tree $\mathsf{Act}~a$ does have a good path, namely $\ChoiceAct$,
but may be unsafe, if the action $a$ is not defined on input state
$s$. For example, the $a$ may be an action for reading from some
pointer $x$, but that pointer may not be allocated in the state $s$.

Safety of a tree will be an important property in the definition of
Hoare triples, where we will require that a precondition of a program
implies that the trees comprising the program's denotation are safe
for every path.

The following are several important lemmas about trees and their
operational semantics, which lift most of the properties of
transitions, to trees.

\begin{lemma}[Coverage of stepping by transitions]
Let $\Delta : V \rightarrow W$, and $\Delta \vdash \bar{x}, s, T
\tsteps{\Path} \bar{x}', s', T'$. Then either the step corresponds to
an idle transition (that is, $(\bar{x}, s) = (\bar{x}', s')$), or
there exists a transition $a \in \intof{V}$, such that
$\unwind~\Delta~a~\bar{x}~s~\bar{x}'~s'$.
\end{lemma}

\begin{lemma}[Other-fixity of stepping]
Let $\Delta : V \rightarrow W$ and $\Delta \vdash \bar{x}, s, T
\tsteps{\Path} \bar{x}', s', T'$. Then $a_o(s) = a_o(s')$.
\end{lemma}

\begin{lemma}[S-preservation of stepping]
Let $\Delta : V \rightarrow W$ and $\Delta \vdash \bar{x}, s, T
\tsteps{\Path} \bar{x}', s', T'$. If $S.W(s)$ then $S.W(s')$.
\end{lemma}

\begin{lemma}[Stability of stepping]
Let $\Delta : V \rightarrow W$ and $\Delta \vdash \bar{x}, s, T
\tsteps{\Path} \bar{x}', s', T'$. Then $s^\top \osteps[W]{} s'^\top$.
\end{lemma}

\begin{lemma}[Determinism of stepping]
Let $\Delta : V \rightarrow W$ and $\Delta \vdash \bar{x}, s, T
\tsteps{\Path} \bar{x}', s', T'$, and $\Delta \vdash \bar{x}, s, T
\tsteps{\Path} \bar{x}'', s'', T''$. Then $\bar{x}' = \bar{x}''$, $s'
= s''$ and $T' = T''$.
\end{lemma}

\begin{lemma}[Locality of stepping]
Let $\Delta : V \rightarrow W$ and $\Delta \vdash \bar{x}, (s \zig p),
T \tsteps{\Path} \bar{x}', s', T'$. Then there exists $s''$ such that
$s' = s''\zig p$, and $\Delta \vdash \bar{x}, (s \zag p), T
\tsteps{\Path} \bar{x}', (s'' \zag p), T'$.
\end{lemma}


\begin{lemma}[Safety monotonicity of stepping]
If $s \zig p \in \tsafe~T~\Path$ then $s \zag p \in \tsafe~T~\Path$.
\end{lemma}

\begin{lemma}[Framability of stepping]
Let $s \zig p \in \tsafe~T~\Path$, and $s \zag p, T \tsteps{\Path} s',
T'$. Then there exists $s''$ such that $s' = s'' \zag p$ and $s \zig
p, T \tsteps{\Path}, s'' \zig p, T'$.
\end{lemma}

The following lemma is of crucial importance, as it relates stepping
with morphisms. In particular, it says that the steps of a tree are
uniquely determined, no matter the morphism under which it
appears. Intuitively, this holds because each transition that a tree
makes has a unique image under a morphism $f : V \rightarrowX X W$.

%

%

\begin{lemma}[Stepping under morphism]
Let $f : V \rightarrowX X W$ and $(s_v, s_w) \in f\ x$. Then the
following hold:
\begin{enumerate}
\item if $s_v, T \tsteps{\Path} s'_v, T'$, then $\exists x'\ s'_w\ldot
  (s'_v, s'_w) \in f\ x'$ and $f \vdash (x), s_w, T \tsteps{\Path}
  (x'), s'_w, T'$.
\item if $f \vdash (x), s_w, T \tsteps{\Path} (x'), s'_w, T'$, then
  $\exists x'\ s'_v\ldot (s'_v, s'_w) \in f\ x'$ and $s_v, T
  \tsteps{\Path} s'_v, T'$.
\end{enumerate}
\end{lemma}

The first property of this lemma relies on the fact that for a step
over states in $V$, we can also find a step over related states in
$W$, i.e., that $f$ encodes a simulation. The second property relies
on the fact that $f$'s state component is a function in the
contravariant direction. Thus, for each $s_w$ there are unique $x$ and
$s_v$, such that $(s_v, s_w) \in f\ x$.

\subsection*{Predicate transformers}
\label{sec:modal-pred}

In this section we define a number of predicate transformers over
trees that ultimately lead to defining the $\vrf$ predicate
transformer on programs.

\begin{definition}
  \label{def:modpred} 
  Let $T : \mathsf{tree}~V~A$, and $\Paths$ be a sequence of
  paths. Also, let $X$ be an assertion over $V$-states and $V$-trees,
  and $Q$ be an assertion over $A$-values and $V$-states. We define
  the following predicate transformers: 
\[
\begin{array}{rcl}
\mathsf{always}^{\Paths}~T~X~s & \eqdef & 
\mathsf{if}~\Paths = \Path :: \Paths' 
~\mathsf{then}~\forall s_2\ldot s \osteps[V]{} s_2 \rightarrow \hbox{}\\
& & \qquad \begin{array}[t]{l}\tsafe~T~\Path~s_2 ~~\wedge X~s_2~T \wedge 
     \forall s_3~T'\ldot~~ s_2, T \tsteps{\Path} s_3, T' \rightarrow
   \mathsf{always}^{\Paths'}~T_2~X~s_3
   \end{array}
\\
&& \mathsf{else}~ \forall s_2. s \osteps[V]{} s_2 \rightarrow X~s_2~T
\\
\mathsf{always}~T~X~s & \eqdef & \forall \Paths\ldot \mathsf{always}^{\Paths}~T~X~s
\\
\\
\mathsf{after}~T~Q & \eqdef & \mathsf{always}~T~(\lambda~s'~T'\ldot\forall v\ldot T' = \mathsf{Ret}~v \implies Q~v~s')
\end{array}
\]

\end{definition}

The helper predicate $\mathsf{always}^{\Paths}~T~X~s$ expresses the
fact that starting from the state $s$, the tree $T$ remains safe and
the user-chosen predicate $X$ holds of all intermediate states and
trees obtained by evaluating $T$ in the state $s$ according to the
sequence of paths $\Paths$. The predicate $X$ remains valid under any
any environment steps of the resource $V$.

The predicate $\mathsf{always}~T~X~s$ quantifiers over the path
sequences. Thus, it expresses that $T$ is safe and $X$ holds after any
finite number of steps which can be taken by $T$ in $s$.

The predicate transformer $\mathsf{after}~T~Q$ encodes that $T$ is
safe for any number of steps; however, $Q\ v\ s'$ only holds if $T$
has been completely reduced to $\mathsf{Ret}\ v$ and state $s'$. In
other words $Q$ is a postcondition for $T$, as it is required to hold
only if, and after, $T$ has terminated.

Now we can define the $\vrf$ predicate transformer on programs, by
quantifying over all trees in the denotation of a program.
\[
\vrf~e~Q~s \eqdef \Sof{V}~s \wedge \forall T\in e\ldot \mathsf{after}~T~Q~s
\]

This immediately gives us a way to define when a program $e$ has a
precondition $P$ and postcondition $Q$: when all the trees in $T$ have
a precondition $P$ and postcondition $Q$ according to the
$\mathsf{after}$ predicate, or equivalently, when
\[
\Sof{V}~s \rightarrow P\ s \rightarrow \vrf~e~Q~s
\]
which is the formulation we used in Section~\ref{sec:formal} to define
the Hoare triples.

We can now state the following soundness theorem, each of whose three
components has been established in the Coq files.

\begin{theorem}[Soundness]~
\begin{itemize}
\item All the properties of $\vrf$ predicate transformer from
  Section~\ref{sec:formal} are valid. 

\item The sets $\mathsf{ST}~V~A$ and $\{P\}\ A\ \{Q\}$ are complete
  lattices under subset ordering with the set
  $\{\mathsf{Unfinished}\}$ as the bottom. Thus one can compute the
  least fixed point of every monotone function by Knaster-Tarski
  theorem.

\item All program constructors are monotone.
\end{itemize}
\end{theorem}

\end{document}